\documentclass[11pt,draftclsnofoot,onecolumn]{IEEEtran}

\usepackage{graphicx,figsize,subfigure}
\usepackage{epsfig}
\usepackage{amsmath}
\usepackage{bbm}
\usepackage{amssymb}
\usepackage{wrapfig}
\usepackage{times,color}
\usepackage{cite,footnote,xspace,syntonly,algorithm,algorithmic,bm}
\usepackage[caption=false,font=footnotesize]{subfig}
\usepackage{ifpdf}
\usepackage{psfrag}
\usepackage{tikz,chemarrow} %for drawing figures
\usepackage{amsfonts}

\input{mysymbol.sty}
\newcommand{\calG}{{\cal G}}

\newcommand{\calN}{{\cal N}}

\def \cN {\mathcal{N}}
\def \cH {\mathcal{H}}

\def \cS {\mathcal{S}}

\def \cP {\mathcal{P}}

\def \bT {{\bf T}}
\def \Ex {{\mathbb{E}}}
\def \bY {{\bf Y}}
\def \bR {{\bf R}}
\def \bS {{\bf S}}
\def \bX {{\bf X}}
\def \bA {{\bf A}}

\def \bW {{\bf W}}

\def \bG {{\bf G}}
\def \bI {{\bf I}}

\def \bB {{\bf B}}
\def \bM {{\bf M}}
\def \bU {{\bf U}}

\def \bE {{\bf E}}
\def \bD {{\bf D}}
\def \bZ {{\bf Z}}
\def \bW {{\bf W}}
\def \bH {{\bf H}}
\def \bL {{\bf L}}
\def \bK {{\bf K}}
\def \bQ {{\bf Q}}

\def \bP {{\bf P}}

\def \bw {{\bf w}}

\def \bx {{\bf x}}
\def \by {{\bf y}}
\def \bz {{\bf z}}
\def \bb {{\bf b}}

\def \bd {{\bf d}}

\def \ba {{\bf a}}
\def \bq {{\bf q}}

\def \chit {{\boldsymbol{\chi}_t}}
\def \bchi {{\boldsymbol{\chi}}}
\def \nut {{\boldsymbol{\nu}_t}}
\def \nust {{\boldsymbol{\nu}^s_t}}
\def \epst {{\boldsymbol{\epsilon}^s_t}}
\def \etat {{\boldsymbol{\eta}_t}}
\def \ceta {{\mathbf{C}_{\boldsymbol{\eta}}}}
\def \cnu {{\mathbf{C}_{\boldsymbol{\nu}}}}
\def \beps {{\boldsymbol{\epsilon}}}

\def \St {{\mathbf{S}_t}}

\long\def\symbolfootnote[#1]#2{\begingroup
\def\thefootnote{\fnsymbol{footnote}}
\footnote[#1]{#2}\endgroup} \psfull

\begin{document}
%--------------------------------------------The First Page---------------------------------------------------------------
% paper title
\title{\huge Dynamic Network Cartography$^\dag$}

\author{{\it Gonzalo~Mateos~(contact author) and Ketan Rajawat$^\ast$}}

\markboth{IEEE SIGNAL PROCESSING MAGAZINE (TO APPEAR)}
\maketitle \maketitle \symbolfootnote[0]{$\dag$ Work in this paper
was supported by the NSF-ECCS grant no. 1202135. The authors would 
like to thank Prof. G. B. Giannakis (U. of Minnesota), for
his invaluable help as PhD advisor.} 
\symbolfootnote[0]{$\ast$ The authors are with the Dept.
of Electrical and Computer Engineering and the Digital Technology Center, U. of
Minnesota, 200 Union Street SE, Minneapolis, MN 55455. Tel/fax:
(612)626-7781/625-4583; Emails:
\texttt{\{mate0058,ketan\}@umn.edu}}

\vspace*{-65pt}
%\begin{center}
%\small{\bf Submitted: }\today\\
%\end{center}
%\vspace*{10pt}

\begin{abstract}
Communication networks have evolved from specialized, research and tactical 
transmission systems to large-scale and highly complex interconnections 
of intelligent devices, increasingly becoming more commercial, consumer-oriented, 
and heterogeneous. Propelled by emergent social networking services
and high-definition streaming platforms, network
traffic has grown explosively thanks to the advances in processing speed and 
storage capacity of state-of-the-art communication technologies.
As ``netizens'' demand a seamless networking 
experience that entails not only higher speeds, but also resilience and 
robustness to failures and malicious cyber-attacks, ample opportunities for signal 
processing (SP) research arise. The vision is for ubiquitous smart network devices 
to enable data-driven statistical learning algorithms for
distributed, robust, and online network operation and 
management, adaptable to the dynamically-evolving network landscape with minimal 
need for human intervention. The present paper aims at delineating 
the analytical background and the relevance of SP tools to dynamic 
network monitoring, introducing the SP readership to the concept of 
\emph{dynamic network cartography} -- a framework 
to construct maps of the dynamic network state in an efficient and
scalable manner tailored to large-scale heterogeneous networks.
\end{abstract}

\vspace*{-25pt}

% % % % % % % % % % % % % % % % % % % % % % % % % % % % % % % % % % % % % % % %
%                         Section I                                           %
% % % % % % % % % % % % % % % % % % % % % % % % % % % % % % % % % % % % % % % %

\section{Introduction}\label{sec:intro}

Emergence of multimedia-enriched social networking services and Internet-friendly portable
devices is multiplying network traffic volume day by
day~\cite{Wu11}. Wireless connectivity under the envisioned dynamic
spectrum paradigm~\cite{tut_rf_cartog} relies on mobile networks of
diverse nodes, which are nevertheless united by unparalleled cognition capabilities,
adaptability, and decision-making attributes. 
Moreover, the advent of networks of intelligent devices such as those deployed to monitor
the smart power grid%~\cite{Ghassemi}
, transportation
networks%~\cite{transportation}
, medical information
networks%~\cite{Phunchongharn}
, and cognitive radio (CR) networks%~\cite{Sadler07}
, will
transform the communication infrastructure to an even more complex
and heterogeneous one. Thus, ensuring compliance to service-level
agreements and quality-of-service (QoS) guarantees 
necessitates breakthrough management and monitoring
tools providing operators with a comprehensive view of the 
network landscape. Situational awareness provided by such tools will
be the key enabler for effective information dissemination, 
routing and congestion control, network health management, risk analysis, and 
security assurance.

But this great promise comes with great challenges. 
Acquiring network-wide performance and utilization metrics 
for large networks is no easy task. Suppose for instance that traffic volumes 
are of interest, not only for gauging instantaneous network health, but also for 
more complex network management tasks such as intrusion detection, capacity provisioning, 
and network planning~\cite{zrwq09}. While traffic volumes on links (also called link counts) are 
readily acquired using off-the-shelf tools such as the simple network management protocol
(SNMP), missing link-count measurements may still skew 
the network operator's perspective. SNMP packets may be dropped for instance, 
if some links become congested, rendering link-count information for those 
links more important, as well as less available~\cite{soule,Roughan}. Classical approaches
relying either on simple time-series interpolation or on regularized least-squares (LS) formulations 
for predicting the missing link counts~\cite{soulekalman}, have not been able to fully 
capture the complexity of the Internet traffic. This is evidenced by the recent upsurge 
of efforts toward advanced network tomography~\cite{tomo_nowak}, and \emph{spatio-temporal} 
traffic estimation algorithms for network monitoring~\cite{soule,zrwq09,pedrocip12}.

Similarly, path metrics such as end-to-end delays 
are of great interest to service providers because they directly affect 
the end-user experience. The challenge here is that the number of paths 
grows very fast as the number of nodes increases. Probing exhaustively 
all origin-destination pairs is impractical and wasteful of resources even for moderate-size 
networks~\cite{nk,Shavitt}. Accurate prediction of missing delays based on the inherent 
e.g., topology-induced correlation 
or smoothness traits among link and path quantities is therefore crucial
for statistical analysis and monitoring tasks~\cite{Kolaczyk_book}.
While the prevailing operational
paradigm adopted in current networks entails nodes continuously communicating
their link measurements to a central monitoring station, 
in-network \emph{distributed} cooperation through local interactions is 
preferred for scalability and robustness considerations~\cite{tsp_rankminimization_2012}.

Conventional network monitoring tools entail a couple of additional limitations. 
First, they are typically resource heavy and tend to overload network 
operators with crude, unrefined data, without enough processing to separate the ``data wheat from the chaff''; 
see e.g., \cite{conti} and references therein. 
It is thus of paramount importance to construct parsimonious descriptors of the \textit{network state}, for the purpose of modeling, monitoring, and management
of complex interconnected systems. 
Due to the diversity of modern networks, the network state can incorporate typical quantities such as traffic volumes and end-to-end delays, as well as latent social metrics such as hierarchy, reputation, and vulnerability. 
Second, malicious activities intended to undermine network functionality or compromise secrecy of data have grown in sophistication, thus rendering traditional
signature-based intrusion detection schemes increasingly obsolete.
Intrusion attempts and malicious attacks manifest themselves as abrupt changes in network states~\cite{Jayasumana}, and such anomalous patterns are oftentimes hidden within the raw high-dimensional network data~\cite{zggr05}. 
For these reasons, unveiling network anomalies in a reliable and computationally-efficient manner is a challenging yet essential goal~\cite{lakhina,tsp_rankminimization_2012,zggr05}.

All in all, accurate network diagnosis and statistical analysis tools are instrumental for maintaining seamless end-user experience in dynamic environments, 
as well as for ensuring network security and stability. 
In this direction, this tutorial advocates the concept of 
\emph{dynamic network cartography} as a tool for statistical modeling, monitoring, and management of complex networks. 
Focus will be placed on two complementary aspects of network cartography, namely, online construction of global 
network state maps using only a few measurements, and unveiling of network anomalies across network flows and time. 
The surveyed cartography algorithms leverage recent advances in machine learning and statistical signal processing 
(SP) methods, including sparsity-cognizant learning, kriged Kalman filtering of dynamical processes over networks,  
nuclear norm minimization for low-rank matrix completion, semi-supervised dictionary learning, and in-network 
optimization via the alternating-directions method of multipliers.
Through a unifying treatment that revolves around network cartography, this paper demonstrates how benefits from foundational SP methods can permeate to dynamic network monitoring, and collectively enable inference of global network health, thus leading to enhanced network robustness and QoS.

% % % % % % % % % % % % % % % % % % % % % % % % % % % % % % % % % % % % % % % %
%                         Section II                                          %
% % % % % % % % % % % % % % % % % % % % % % % % % % % % % % % % % % % % % % % %

\section{Global performance prediction via dynamical network cartography}
\label{sec:dnc}

This section deals with the problem of \emph{mapping the network state} from incomplete 
sets of measurements, and touches upon two application domains. 
A dictionary learning algorithm is introduced first to efficiently 
impute  missing link traffic volumes, using measurements from a 
wide class of (possibly non-stationary) traffic patterns~\cite{pedrocip12}.
Subsequently, the problem of tracking and predicting end-to-end network delay
is considered, and the dynamic network kriging approach of~\cite{rajassp12} is described. 

% % % % % % % % % % % % % % % % % % % % % % % % % % % % % % % % % % % % % % % %
%                         Subsection II-A                                     %
% % % % % % % % % % % % % % % % % % % % % % % % % % % % % % % % % % % % % % % %

\subsection{Semi-supervised dictionary learning for traffic maps}
\label{ssec:ssdl}
%Traffic volume measurements are often replete with missing data. 
%A widely studied problem in this context is that of estimating origin-destination (OD)
%traffic matrices, with most techniques relying on stationarity assumptions regarding traffic 
%volumes~\cite{Kolaczyk_book,zrwq09}. This section describes a recent 
%approach that utilizes dictionary learning for imputing missing link traffic volumes, and is shown 
%to efficiently handle a much wider class of non-stationary traffic patterns~\cite{pedrocip12}.

Consider an Internet protocol (IP) network comprising $N$ nodes and $L$ links,
carrying the traffic of $F$ origin-destination flows (network connections).
Let $x_{l,t}$ denote 
the traffic volume (in bytes or packets) passing through link $l \in \{1,\ldots,L\}$ over a fixed 
interval of time $(t,t+\Delta t)$. Link counts across the entire network 
are collected in the vector $\bx_t\in\mathbb{R}^L$, e.g., using the ubiquitous SNMP protocol. 
Since measured link counts are both unreliable and incomplete due to hardware or software 
malfunctioning, jitter, and communication errors~\cite{zrwq09,Roughan}, 
they are expressed as noisy versions of a subset of $S<L$ links
\begin{equation} \label{meas}
\by_t = \bS_t \bx_t + \beps_t,\quad t=1,2,\ldots
\end{equation}
where $\bS_t$ is an $S \times L$ selection matrix with 0-1 entries whose 
rows correspond to rows of the identity matrix of size $L$, and $\beps_t$ 
is an $S \times 1$ zero-mean noise term with constant variance 
accounting for measurement and synchronization errors. Given $\by_t$ the aim is 
to form an estimate $\hat{\bx}_t$ of the full vector of link counts $\bx_t$, 
which in this case defines the network state.

A simple approach implemented in 
measurement-processing software such as RRDtool~\cite{rrd}, 
is to ignore the noise term and rely on one-dimensional interpolation for 
the time series $\{x_{l,t}\}$ per link $l$. The applicability and accuracy of this 
scheme is however limited, since it tacitly assumes that the entries 
of $\bx_t$ are uncorrelated; missing entries $x_{l,t}$ are few 
and do not occur in bursts; and the time series $\{\bx_t\}$ is stationary. 
Nevertheless, none of these assumptions holds true in real networks~\cite{Roughan}. 

The reliance on stationarity and availability of measurements 
from contiguous time intervals can be forgone if estimation of ${\bx_t}$ 
is performed for each $t$ individually. 
In principle, $\hat{\bx}_t$ can be obtained if the volumes of origin-destination (OD) 
traffic flows $\bz_t\in\mathbb{R}^F$ are available, since they are related through
% by utilizing the following relationship
%
\begin{align}\label{eq:ltra}
\bx_t = \bR \bz_t 
\end{align}
where the so-termed routing matrix $\bR:=[r_{l,f}]\in\{0,1\}^{L\times F}$ 
is such that $r_{l,f} = 1$ if link $l$ carries the flow $f$, and 
zero otherwise. However, measuring $\bz_t$ is even more difficult and in practice 
$\bz_t$ is itself estimated from $\{\bx_t\}$ through \textit{tomographic traffic 
inference}~\cite{tomo_nowak,Kolaczyk_book}, 
where given $\bR$ and noisy link counts, the goal is to estimate the OD flows as the solution of a  
linear inverse problem.
% and the inverse problem of inferring $\bz_t$ from $\{\bx_t\}$ has been extensively studied over the past 10 years. 
Since the inverse problem is highly under-determined 
$\left[F=\mathcal{O}(N^2)\gg L=\mathcal{O}(N)\right]$, early approaches 
relied on prior knowledge in the form of statistical models for the OD flows 
(such as the Poisson, Gaussian, logit-choice, or gravity models), that ultimately 
serve as complexity-controlling (that is regularization) mechanisms~\cite[Ch. 9]{Kolaczyk_book}.
Among these, the state-of-the-art traffic matrix estimation algorithm uses an 
entropy-based regularizer, and has been shown to be fast, accurate, robust, 
and flexible \cite{Zha05est}. %Finally, within the context of traffic matrix estimation, 
Time-series analysis-based approaches (such as the Kalman filter in~\cite{soulekalman}) have also been 
proposed for scenarios where link-count measurements are available over contiguous time slots.

Recently, a link-count prediction algorithm was put forth in~\cite{pedrocip12}, 
where missing entries of $\bx_t$ are estimated from 
historical measurements in $\mathcal{T}_{S}:=\{{\by}_t\}_{t=1}^T$ by leveraging
the structural regularity of $\bR$ through a semi-supervised dictionary learning 
(DL) approach. Under the DL framework, \emph{data-driven} dictionaries for \emph{sparse} signal 
representation are adopted as a versatile means of capturing parsimonious signal
structures; see e.g.,~\cite{Dictionary_learning_SP_mag_10} for a tutorial treatment. 
Propelled by the success of compressive sampling (CS)~\cite{Do06CS}, sparse signal modeling has 
led to major advances in several machine learning, audio and image processing 
tasks~\cite{Dictionary_learning_SP_mag_10,Ti1996lasso}. 
Motivated by these ideas, it is postulated in~\cite{pedrocip12} that link counts can 
be represented as a linear combination $\bx_t = \bB\bw_t$ of a 
few ($\ll Q$) columns of an over-complete dictionary (basis) matrix 
$\bB:=[\bb_1,\ldots,\bb_Q]\in\mathbb{R}^{L \times Q}$, 
where $\bw_t\in\mathbb{R}^Q$ is a \textit{sparse} vector of expansion coefficients. 
Many signals including speech and natural images admit sparse representations even under 
generic predefined dictionaries, such as those based on the Fourier and the wavelet 
bases, respectively~\cite{Dictionary_learning_SP_mag_10}. 
Like audio and natural images, link counts can exhibit strong 
correlations as evidenced from the structure of $\bR$ [cf. \eqref{eq:ltra}]. For instance, the traffic 
volumes on links $i$ and $j$ are highly correlated if they both carry common flows.
DL schemes are attractive due to their flexibility, since 
they utilize training data to \textit{learn} an appropriate over-complete basis customized 
for the data at hand.
However, the use of DL for modeling network data is well motivated but so far
relatively unexplored.  

\noindent\textbf{Prediction of link counts.} Suppose for now that either a learnt, or,
a suitable pre-specified dictionary $\bB$ is available, and consider
predicting the missing link counts. Data-driven learning of dictionaries
from historical data will be addressed in the ensuing subsection.
Given $\bR$ and the link count measurements $\by_t$, contemporary tools developed in the area of CS 
and semi-supervised learning can be used to form $\hat{\bx}_t$, 
which includes estimates for the missing $L-S$ 
link counts~\cite{Ti1996lasso,Do06CS,Be2006Mreg}. The spatial regularity of 
the link counts is captured through the auxiliary weighted graph $\calG$ with $L$ vertices, 
one for each link in the network. The edge weights for all edges in $\calG$ 
are subsumed by the off-diagonal entries of the Gram matrix $\bG=[g_{i,j}]:=
\bR\bR'\in\mathbb{R}^{L\times L}$,
where $(\cdot)'$ denotes transposition. The off-diagonal 
entries $g_{i,j}$ count the  number of OD flows that are common to both 
links $i$ and $j$. Main diagonal entries of $\bG$ count the number of OD 
flows that use the corresponding links.

Given a snapshot of incomplete link counts $\by_t$ during the 
\emph{operational phase} (where a suitable basis $\bB$ is available), the sparse 
basis expansion coefficient vector $\bw_t$ is estimated as
\begin{equation}\label{eq:opphase}
\hat{\bw_t}:=\mathop{\arg\min}_{\bw_t}\|\by_t-\bS_t\bB\bw_t\|_2^2
+\lambda_w\|\bw_t\|_1+\lambda_g\bw_t'\bB'\bL\bB\bw_t
\end{equation} 
where $\bL:=\textrm{diag}(\bG\mathbf{1}_L)-\bG$ denotes the Laplacian 
matrix of $\calG$; $\lambda_w,\lambda_g>0$ are 
tunable regularization parameters; and $\mathbf{1}_L$ 
is the $L\times 1$ vector of all ones. 
The criterion in \eqref{eq:opphase} consists of
a LS error between the observed and postulated link counts, 
along with two regularizers. The $\ell_1$-norm $\|\bw_t\|_1$ 
encourages sparsity in the coefficient vector $\hat\bw_t$~\cite{Do06CS,Ti1996lasso}. 
With $\bx_t:=[x_{1,t},\ldots,x_{L,t}]'$ given by $\bx_t = \bB\bw_t$, 
the Laplacian regularization can be explicitly written as
$\bw_t'\bB'\bL\bB\bw_t=(1/2)\sum_{i=1}^L\sum_{j=1}^Lg_{i,j}(x_{i,t}-x_{j,t})^2.$
It is thus apparent that $\bw_t'\bB'\bL\bB\bw_t$
encourages the link counts to be close if their corresponding vertices
are connected in $\calG$. Each summand is weighted 
according to the number of OD flows common to links $i$ and $j$. 
Typically adopted for semi-supervised learning, such a regularization 
term encourages $\bB\bw_t$ to lie on a smooth 
manifold approximated by $\calG$, which constrains how the measured link counts relate to $\bx_t$ 
\cite{Be2006Mreg,Ra2007SLT}. It is also common to use normalized 
variants of the Laplacian instead of $\bL$~\cite[p. 46]{Kolaczyk_book}.

The cost in \eqref{eq:opphase} is convex but non-smooth, 
and customized solvers developed for $\ell_1$-norm regularized 
optimization can be employed here as well, e.g.,~\cite{Fri07pwise}. 
Once $\hat{\bw}_t$ is available, an estimate of the full vector of link counts 
is readily obtained as $\hat{\bx}_t:=\bB\hat{\bw}_t$. It is apparent that 
the quality of the imputation depends on the chosen $\bB$, and DL from
historical network data in $\mathcal{T}_{S}$ is described next.

\begin{figure}
\centering
%drawing for propagation environment
\begin{tikzpicture}[auto,scale=1.3]
	\tikzstyle{block} = [draw,rectangle,thick,minimum height=2em,minimum width=2em, line width =1pt];
	\tikzstyle{connector} = [->,thick];
	\tikzstyle{connector} = [->,thick];
	\tikzstyle{branchb} = [circle,inner sep=0pt,minimum size=0.2mm,fill=blue,draw=blue];
	\tikzstyle{branchr} = [circle,inner sep=0pt,minimum size=0.2mm,fill=red,draw=red];
	\tikzstyle{legend_training}=[rectangle, rounded corners, thin,
   top color= white,bottom color=blue!25,                    minimum width=2.5cm, minimum height=0.8cm,violet];
   \tikzstyle{legend_op}=[rectangle, rounded corners, thin,
   top color= white,bottom color=red!25,                    minimum width=2.5cm, minimum height=0.8cm,red];
	
	%draw grid
	%\draw[step=.1cm, gray, very thin] (-2.0,-1.6) grid (6,0.2);    
	% node placement with matrix library: 5x4 array
    \matrix[ampersand replacement=\&, row sep=0.5cm, column sep=0.2cm] {
   \node[branchb] (l0){};
      \&
      \&
       \&
       \&
      \\
      
      \node[block] (T1) {\small $\displaystyle\min_{\mathbf{w}_t} C_t (\bB[k],\mathbf{w}_t)$}; 
      \&
       \&
      \&
      \node[branchr] (l1) {};
       \& \\
\node[block] (T2) {\small$\displaystyle\min_{\|\bb_q\| \leq 1}\displaystyle\sum_{t=1}^T 
C_t(\bB,\mathbf{w}_t[k])$};
      \&
      \&
      \&
      \node[block] (O1) {\small $\displaystyle\min_{\mathbf{w}_t} 
      C_t(\hat{\mathbf{B}},\mathbf{w}_t)$};\&
       \\

      \&
      \&
       \&
      \node[block] (O2) {\small$\hat{\mathbf{x}}_t=\hat{\mathbf{B}}\hat{\mathbf{w}}_t$};
      \&
      \\
      
      \&
      \&
       \&
       \node[branchr] (l2){};
      \&
      \\
    };
    %draw lines
    \draw [connector,color=blue,line width= 1pt] (T1) -- node {${\mathbf{W}}_t[k]$} (T2);
    \draw [connector,color=red] (O1) -- node {$\hat{\mathbf{w}}_t$} (O2);
    \draw [connector,line width= 1pt] (T2) -- node {$\hat{\mathbf{B}}$} (O1);
    \draw [connector,color=blue,line width= 1pt] (T2.south) -- ++(0,-.5cm) -|node [near 
start]{$\bB[k+1]$} ++(-1.8cm,0cm) -| ++(0cm,2.45cm) -> ++ (0.7cm,0cm);
    \draw [connector,color=red,line width= 1pt] (l1.south) -> node [near start]{$\mathbf{y}_t,\: t> T$}(O1);
    \draw [connector,color=blue,line width= 1pt] (l0.south) -> node [near start]{$\{\mathbf{y}_t\}_{t=1}^T$}(T1.north);
    \draw [connector,color=red,line width= 1pt] (O2.south) -> node [near end]{$\hat{\mathbf{x}}_t$}(l2.north);
  %%%%%%%%%%%%
  %insert label for operational and training phase
  \node[legend_training] at (-1.5,2.7){\textsc{\small Training Phase}};
  \node[legend_op] at (2.0,2.7){\textsc{\small Operational Phase}};
	\draw[line width=1pt,color=green!50!black,dashed] (0.5,2.3) --(0.5,-2);  
\end{tikzpicture}
\vspace{-0.4cm}
\caption{Training and operational phases of the semi-supervised DL approach for
link-traffic cartography in~\cite{pedrocip12}.}
\label{fig:ssdlblk}
\vspace{-0.8cm}
\end{figure}
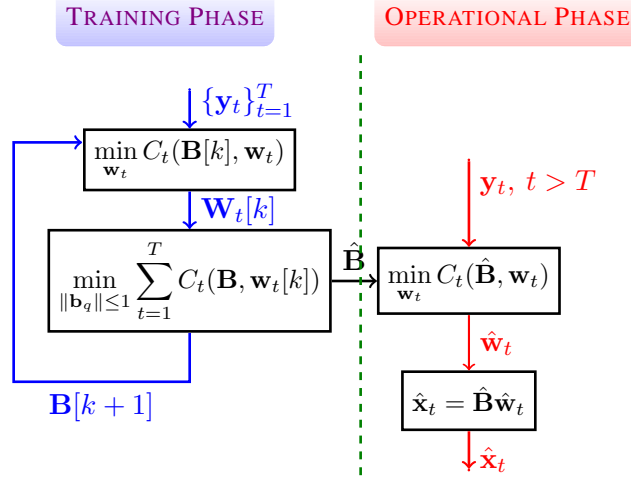

\noindent\textbf{Data-driven dictionary learning.} In its canonical form, 
DL seeks a (typically fat) dictionary $\bB$ so that training data 
$\mathcal{T}_L:=\{\bx_t\}_{t=1}^T$ are well approximated as 
$\bx_t\approx\bB\bw_t$, $t=1,\ldots,T$, for some sparse vectors $\bw_t$ of expansion
coefficients~\cite{Dictionary_learning_SP_mag_10}. Standard DL algorithms cannot, however, 
be directly applied to learn $\bB$ since they rely on the entire vector $\bx_t$. 
To learn the dictionary in the \emph{training phase} using incomplete
link counts $\mathcal{T}_{S}$ instead of $\mathcal{T}_{L}$, the idea is to 
capitalize on the structure in $\bx_t$, of which
$\calG$ is an abstraction~\cite{pedrocip12}. To this end, one can adopt a 
similar cost function as in the operational phase [cf. \eqref{eq:opphase}], 
yielding the data-driven basis and the corresponding sparse representation
\begin{equation}\label{eq:ssdl}
\{\hat{\bW},\hat{\bB}\}:=\mathop{\arg\min}_{\bW,\bB:\{\|\bb_q\|_2\leq 1\}_{q=1}^Q}
\sum_{t=1}^T \left[\|{\by}_t-\bS_t\bB\bw_t\|_2^2\!+\!\lambda_w\|\bw_t\|_1\!+
\!\lambda_g\bw_t'\bB'\bL\bB\bw_t\right]
\end{equation}
where $\hat\bW:=[\hat{\bw}_1,\ldots,\hat{\bw}_T]\in\mathbb{R}^{Q\times T}$. 
The constraints $\{\|\bm\bb_q\|_2\leq 1\}_{q=1}^Q$ 
remove the scaling ambiguity in the products $\bB\bw_t$, and prevent the 
entries in $\bB$ from growing unbounded. Again, the combined regularization 
terms in \eqref{eq:ssdl} promote both sparsity in $\bw_t$ through the $\ell_1$-norm, 
and smoothness across the entries of $\bB\bw_t$ via the 
Laplacian $\bL$. The regularization parameters $\lambda_w$ and
$\lambda_g$ are typically cross-validated~\cite{Ti1996lasso,Fri07pwise}. 
Although \eqref{eq:ssdl} is non-convex, a block coordinate-descent (BCD) solver still 
guarantees convergence to a stationary point~\cite{Bertsekas_Book_Distr}. 
The BCD updates involve solving for $\bB$ and $\bW$ 
in an alternating fashion, both doable efficiently via convex programming~\cite{pedrocip12}.
Alternatively, the online DL algorithm in~\cite{mairal} offers enhanced 
scalability by sequentially processing the data in $\mathcal{T}_S$. 
The training and operational (prediction) phases are summarized in Fig. \ref{fig:ssdlblk}, where 
$C_t(\bB,\mathbf{w})$ denotes the $t$-th summand from the 
cost in \eqref{eq:ssdl}.

\begin{figure}
\centering
%\psfrag{MSE}{\footnotesize RE ($\times 10^5$)}
\includegraphics[width=0.6\linewidth]{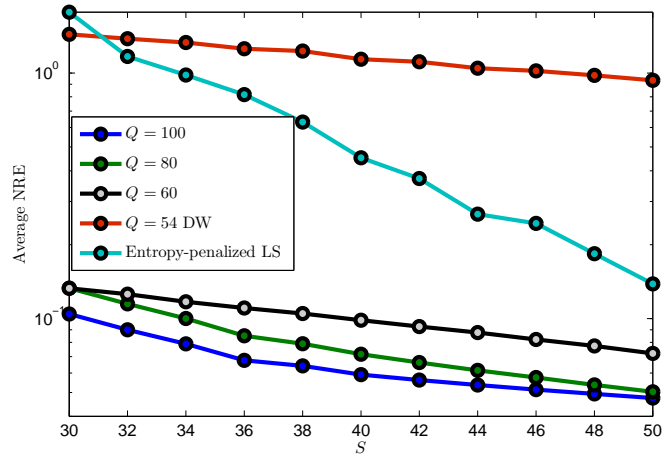}
\vspace{-0.4cm}
\caption{Link-traffic cartography of Internet2 data~\cite{Internet2}. 
Comparison of NRE for different values of $S$~\cite{pedrocip12}.}
\label{varyM50}
\vspace{-0.8cm}
\end{figure}

The explicit need for Laplacian regularization is apparent from \eqref{eq:ssdl}. 
Indeed, if measurements from a certain link are not present in $\mathcal{T}_{S}$, 
the corresponding row of $\bB$ may still be estimated with reasonable 
accuracy because of the third term in $C_t(\bB,\mathbf{w})$. On top of that,
it is because of Laplacian regularization that the prediction performance
degrades gracefully as the number of missing entries in $\by_t$ increases; see also Fig. \ref{varyM50}. 
It is worth stressing that the time series $\{\by_t\}$ need not be 
stationary or even contiguous in time. The link-traffic cartography approach 
described so far can also be adapted to
accommodate time-varying network topologies or routing matrices, 
using a time-dependent Laplacian $\bL_t$. A word of caution is due however, 
since drastic changes in either $\bL_t$ or in the statistical properties of the
underlying OD flows $\bz_t$, 
will necessitate re-training $\bB$ to attain satisfactory performance. 
Finally, note that DL techniques
incur a complexity at least cubic in the size of the network, and 
are better suited for monitoring of backbone wide-area networks which are typically not very large.

Next, a numerical test on link count data from the Internet2 
measurement archive \cite{Internet2} is outlined. The data consists of link counts, 
sampled at 5 minute intervals, collected over several weeks. For the purposes of 
comparison, the training phase consisted of 2000 time slots, with a random subset 
of $50$ links measured (out of $L = 54$ per time slot. The performance of the learned 
dictionary is then assessed over the next $T_0=2000$ time slots. 
Each test vector $\by_t$ is constructed by randomly selecting $S$ entries of 
the full link count vector $\bx_t$. The tuning parameters are chosen via 
cross-validation ($\lambda_s = 0.1$ and $\lambda_g = 10^{-5}$). Fig.~\ref{varyM50} 
shows the normalized reconstruction error (NRE), evaluated as 
$(LT_0)^{-1}\sum_{t=1}^{T_0}\left\|\by_t-\hat{\bx}_t\right\|^2$ for different values 
of $Q$ and $S$. For comparison, the prediction performance with a fixed diffusion wavelet 
matrix \cite{coates} (instead of the data-trained dictionary), as well as that of the 
entropy-penalized LS method \cite{Zha05est} is also shown. The latter approach solves 
a LS problem augmented with a specific entropy-based regularizer, that encourages the 
traffic volumes at the source/destination pairs to be stochastically independent. The 
DL-based method markedly outperforms the competing approaches, especially for low 
values of $S$. Furthermore, note how performance degrades gracefully as $S$ decreases.
Remarkably, the predictions are close to the actual traffic even when using only 30
link counts during the prediction phase.

% % % % % % % % % % % % % % % % % % % % % % % % % % % % % % % % % % % % % % % %
%                         Subsection II-B                                     %
% % % % % % % % % % % % % % % % % % % % % % % % % % % % % % % % % % % % % % % %

\subsection{Delay cartography via dynamic network kriging}
\label{ssec:dnk}
Instead of link counts, consider now the problem of monitoring delays $d_{p,t}$ on a set of 
multihop paths $p \in \cP$, that connect $P:=|\cP|$ source-destination pairs in an IP network.
Path delays are important metrics required by network 
operators for assessment, planning, and fault diagnosis~\cite{Kolaczyk_book,nk,rajassp12}. 
However, monitoring path metrics 
is challenging primarily because $P$ generally grows as the square of the
number of nodes in the network. 
Therefore, at any time $t$ delays can only be measured on a subset of 
paths $\cS_t \subset \cP$, collected in the vector $\bd^s_t$. 
Based on the partial current and past measurements 
$\cH_t:=\{\bd^s_{\tau}\}_{\tau=1}^t$, \emph{delay cartography} amounts to predicting
the remaining path delays $\bd^{\bar{s}}_t:=\{d_{p,t}\}_{p \in \cP \setminus \cS}$. 

A promising approach in this
context has been the application of \emph{kriging}, a tool for 
spatial prediction popular in geostatistics and environmental sciences~\cite{cressie}.
A \emph{network} kriging scheme was developed in~\cite{nk}, which advocates prediction 
of network-wide path delays using measurements on a fixed subset of paths.
The class of linear predictors introduced therein leverages network topology information to model
the covariance among path delays. Building on these ideas, a 
\emph{dynamic} network kriging approach capable of real-time spatio-temporal delay 
predictions was put forth in~\cite{rajassp12}. Specifically, 
a kriged Kalman filter is employed to explicitly capture temporal 
variations due to queuing delays, while retaining the topology-based 
spatial kriging predictor. 
The per-path delay $d_{p,t}$ comprises several independent components 
due to contributions from each intermediate link and router, 
and is modeled in~\cite{rajassp12} as 
\begin{align}
d_{p,t} = \chi_{p,t} + \nu_{p,t} + \epsilon_{p,t}. \label{pathcomp} 
\end{align}
The queuing delay $\chi_{p,t}$ (collected in $\chit\in\mathbb{R}^{P}$)
depends on the traffic, and exhibits spatio-temporal correlation, periodic behavior as well 
as occasional bursts, prompting the following random walk model
\begin{align}
\chit = \bchi_{t-1} + \etat \label{randwalk}
\end{align}
where the driving noise $\etat$ has zero mean and covariance matrix $\ceta$. 
The second term in~\eqref{pathcomp}, collected in the vector 
$\nut$, combines the processing, transmission, and propagation delays, 
and is temporally white but spatially 
correlated, owing to the overlap between paths. Similar to~\cite{nk}, 
the correlation between two paths is modeled as being proportional to the number of links they share, 
so that the covariance matrix $\cnu = \alpha \bU \bU'$, where, 
$u_{p,l}=1$ if path $p$ contains link $l$, and $u_{p,l}=0$ otherwise. 
Finally, the noise term $\epsilon_{p,t}$ is zero mean i.i.d. 
with known variance $\sigma^2$. Defining the $S 
\times P$ path selection matrix as in Sec. \ref{ssec:ssdl}, the measurement 
equation can be written as (introduce $\nust:=\St\nut$ and likewise $\epst$)
\begin{align}
\bd^s_t = \St\chit + \nust + \epst. \label{measeqn}
\end{align}

\begin{figure}[ht]
\begin{minipage}[b]{0.48\linewidth}
  \centering
  \centerline{\includegraphics[width=\linewidth, height=2 in]{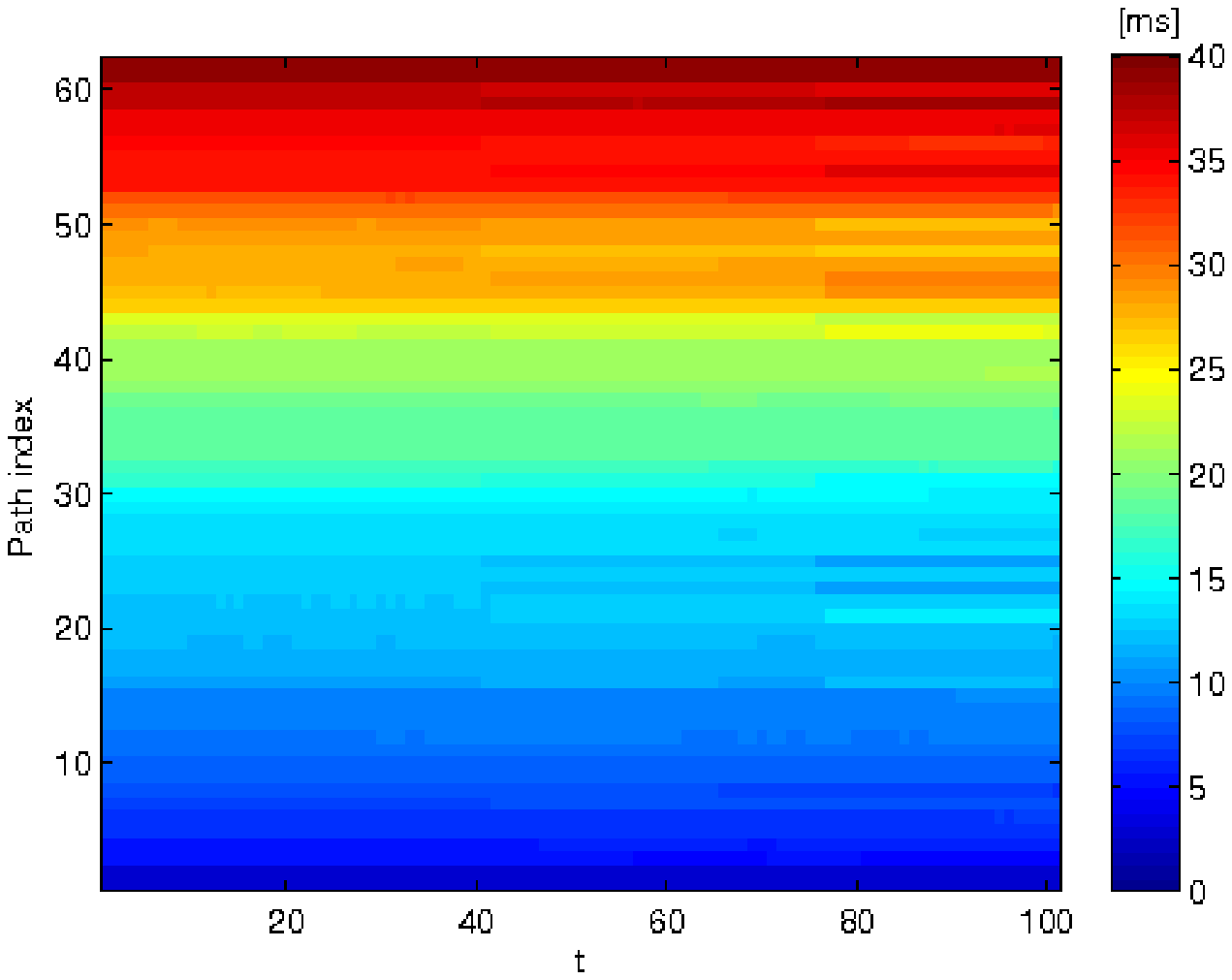}}
\medskip
\end{minipage}
\hfill
\begin{minipage}[b]{.48\linewidth}
  \centering
  \centerline{\includegraphics[width=\linewidth, height=2 in]{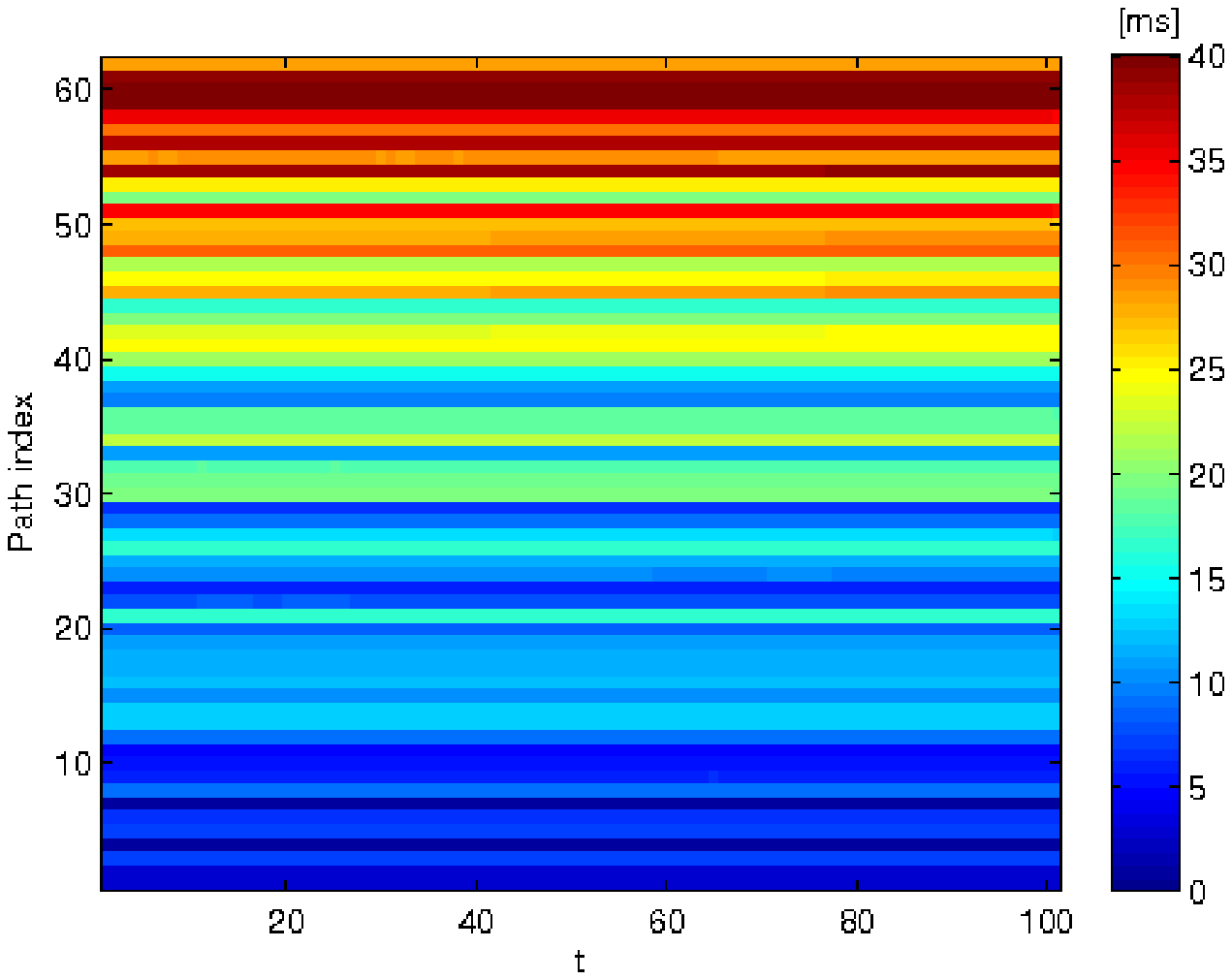}}
\medskip
\end{minipage}\\
\begin{minipage}[b]{0.48\linewidth}
  \centering
  \centerline{\includegraphics[width=\linewidth, height=2 in]{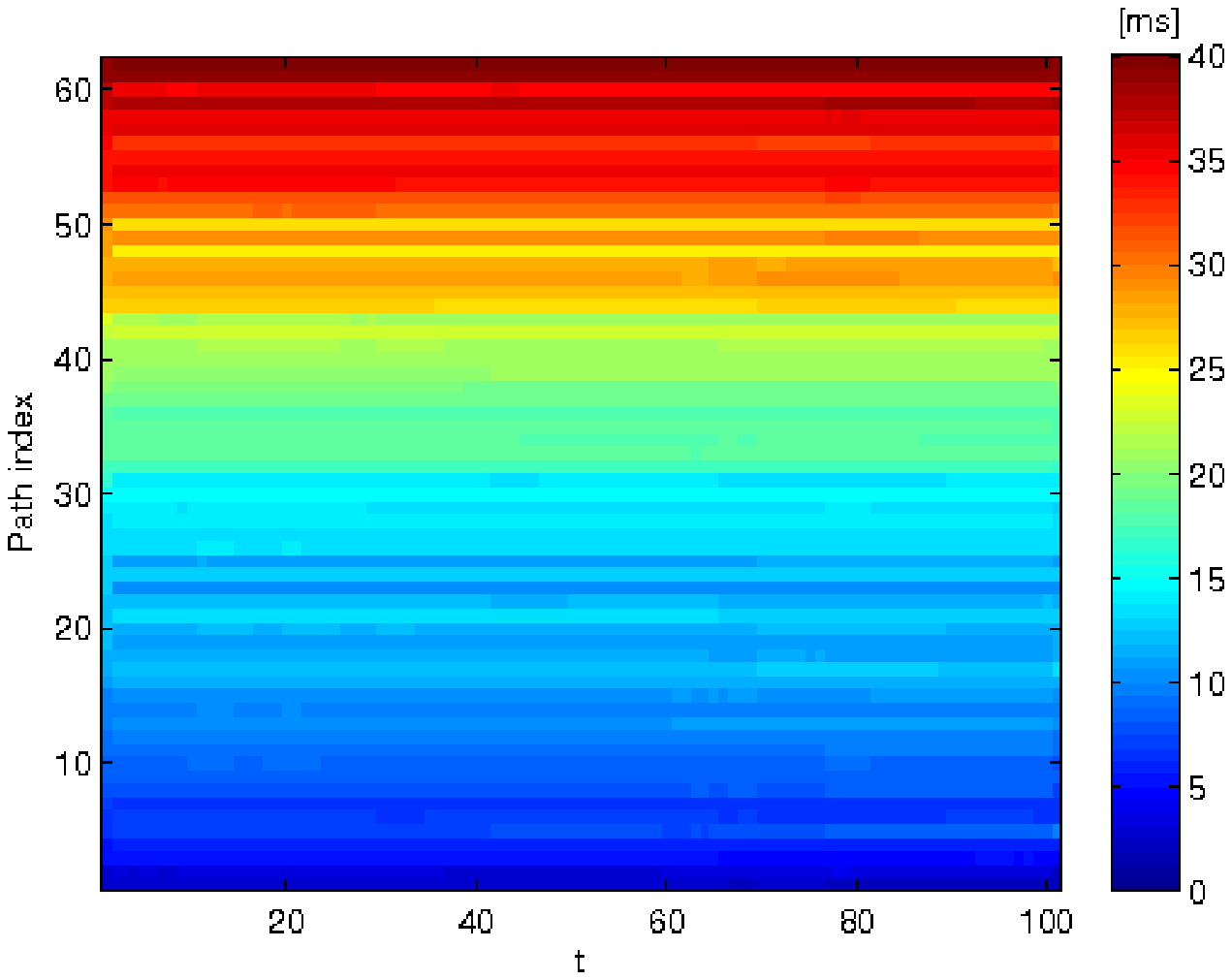}}
\medskip
\end{minipage}
\hfill
\begin{minipage}[b]{.48\linewidth}
  \centering
  \centerline{\includegraphics[width=\linewidth, height=2 in]{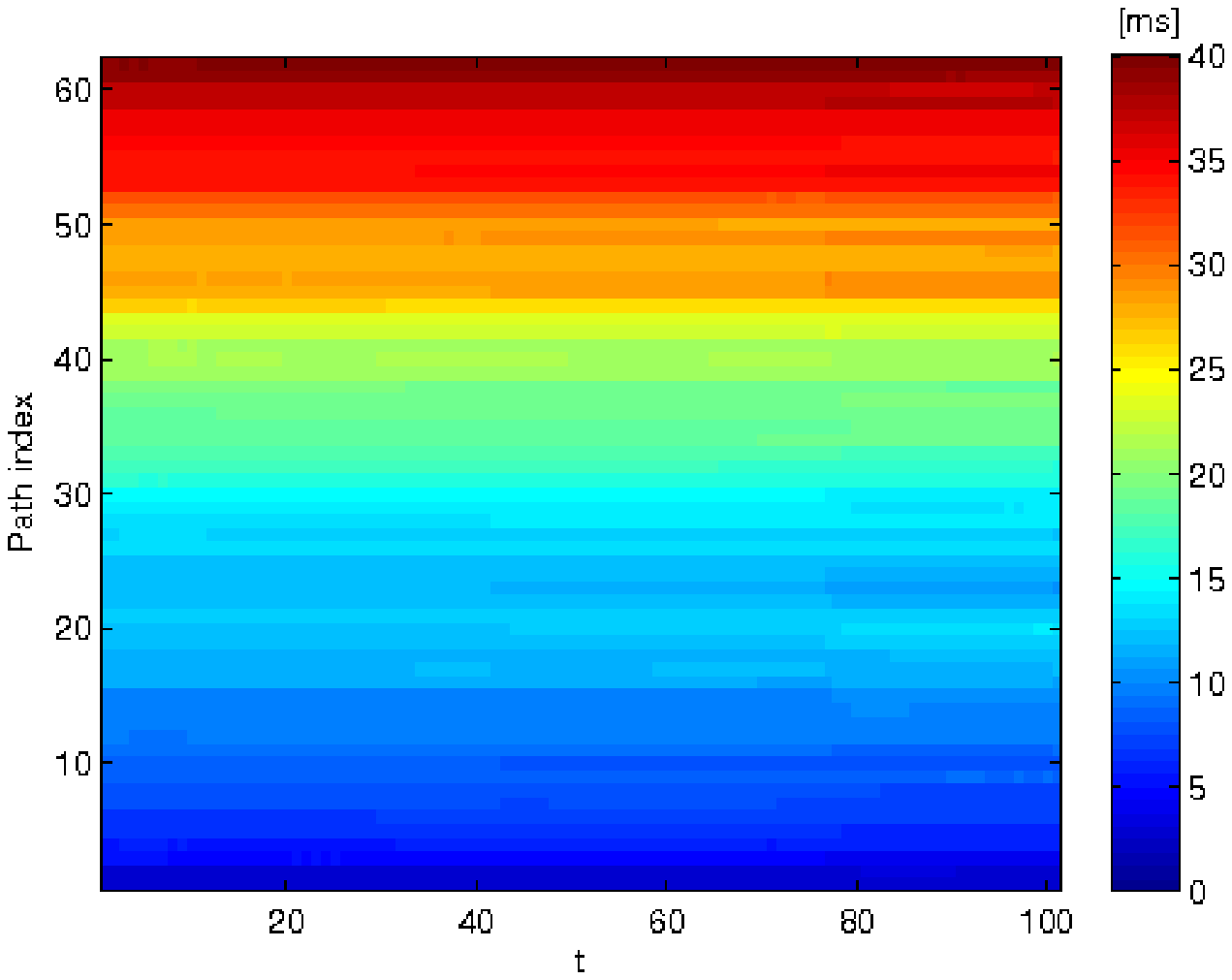}}
\medskip
\end{minipage}
\vspace{-0.4cm}
\caption{True and predicted delay map for $62$ paths in the Internet-2 dataset~\cite{Internet2} 
over an interval of $100$ minutes. 
(Top-left) True delays; (Top-right) network kriging~\cite{nk}; 
(Bottom-left) difussion wavelets~\cite{coates}; and (Bottom-right)
KKF~\cite{rajassp12}. Delays of several paths change
slightly around $t = 80$, but this change is only discernible
from the delay predictions offered by KKF. Delay maps summarize the network state,
and are useful tools aiding
operational decision in network monitoring and control stations~\cite{rajassp12}.}
\vspace{-0.8cm}\label{fig:delay_cart} 
\end{figure}

In the absence of $\St$, the spatio-temporal model in \eqref{randwalk}-\eqref{measeqn} 
is widely employed in geostatistics, where $\chit$ is generally referred to as 
trend, and $\nut$ captures the random fluctuations around $\chit$; see e.g.~\cite{MGRA98}.
Similar models have been employed in~\cite{kim} to describe the dynamics of wireless propagation 
channels, and in~\cite{Cor09} for spatio-temporal random field estimation. 
For a static selection matrix, i.e., $\St:=\bS$ for 
all $t$, the network kriging approach~\cite{nk} entails the following 
two-step procedure: (s1) treat $\nust$ as noise, and 
estimate $\chit$ using the generalized LS criterion; and  (s2) 
use the aforesaid estimate to find the linear minimum mean-square error (LMMSE) 
estimator (denoted by $\Ex^{*}$) for $\nust$, namely
\begin{align}
\Ex^{*}\left[\nust | \chit\right] = \bS \cnu \bS' \left(\bS\cnu\bS' + \sigma^2\bI_S\right)^{-1}
\left[\bd^s_t - \St\chit\right].
\end{align}
Recently, a CS-based approach has also been reported 
for predicting network-wide performance metrics~\cite{coates}. For instance, 
diffusion wavelets were utilized in \cite{coates} to obtain a compressible 
representation of the delays, and account for spatial \emph{and} temporal 
correlations. Although this allows for enhanced prediction accuracy relative to~\cite{nk}, 
it requires batch processing of measurements which does not scale well to 
large networks for real-time operation. Pictorially, the performance of different 
algorithms can be assessed through the delay maps shown in Fig. \ref{fig:delay_cart}.

The spatio-temporal model set forth earlier can provide a better estimate 
of $\chit$ by efficiently processing both present and past measurements jointly. 
Towards this end, a Kalman filter is employed in~\cite{rajassp12}, 
which at time $t$ yields the following update equations
\begin{align}
\hat{\bchi}_t &:= \mathbb{E}^{*}\left[\chit | \mathcal{H}_t\right] = 
\hat{\bchi}_{t-1} + \bK_t (\bd^s_t - \St 
\hat{\bchi}_{t-1} ) \nonumber\\
\bM_t &:= \mathbb{E} \left[(\chit - \hat{\bchi}_t)(\chit-\hat{\bchi}_t)' \right] = 
(\bI_P - \bK_t\St)(\bM_{t-1}+\cnu)\nonumber
\end{align}
where $\bK_t := (\bM_{t-1} + \cnu)\bS'_t \left[\St(\cnu + \ceta + \bM_{t-1})
\bS'_t+\sigma^2\bI_S\right]^{-1}$ is the so-termed Kalman gain. 
The final predictor, referred also as the \textit{kriged} Kalman filter (KKF), is given by
\begin{align}
\hat{\bd}^{\bar{s}}_t := \bar{\bS}_t\hat{\bchi}_t + \bar{\bS}_t\cnu\bS'_t\left(\St\cnu\bS'_t + 
\sigma^2\bI_S\right)^{-1}[\bd^s_t - \St\hat{\bchi}_t]\nonumber
\end{align}
and the prediction error covariance matrix is
\begin{equation*}
\bM^{\bar{s}}_t := \mathbb{E}\left[\left(\bd^{\bar{s}}_t - \hat{\bd}^{\bar{s}}_t\right)
\left(\bd^{\bar{s}}_t - \hat{\bd}^{\bar{s}}_t\right)'\right]
= \sigma^2\bI_S + \bar{\bS}_t\left[\left(\bM_{t-1} + \cnu + \ceta \right)^{-1} + 
\frac{1}{\sigma^2}\bS'_t\St\right]^{-1}\bar{\bS}'_t.
\end{equation*}

\begin{figure}[ht]
\begin{minipage}[b]{0.48\linewidth}
  \centering
  \centerline{\includegraphics[width=\linewidth, height=2 in]{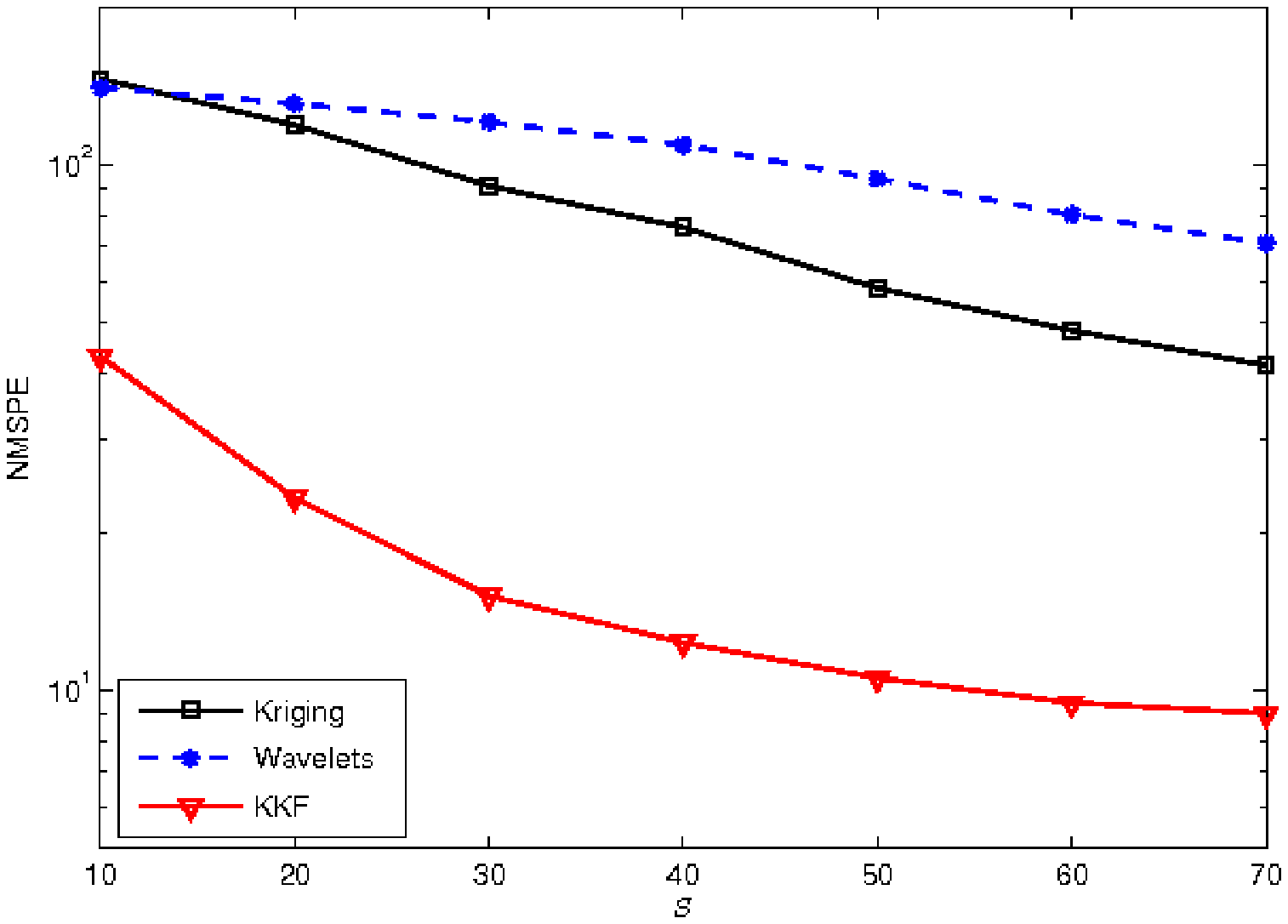}}
\medskip
\end{minipage}
\hfill
\begin{minipage}[b]{.48\linewidth}
  \centering
  \centerline{\includegraphics[width=\linewidth, height=2 in]{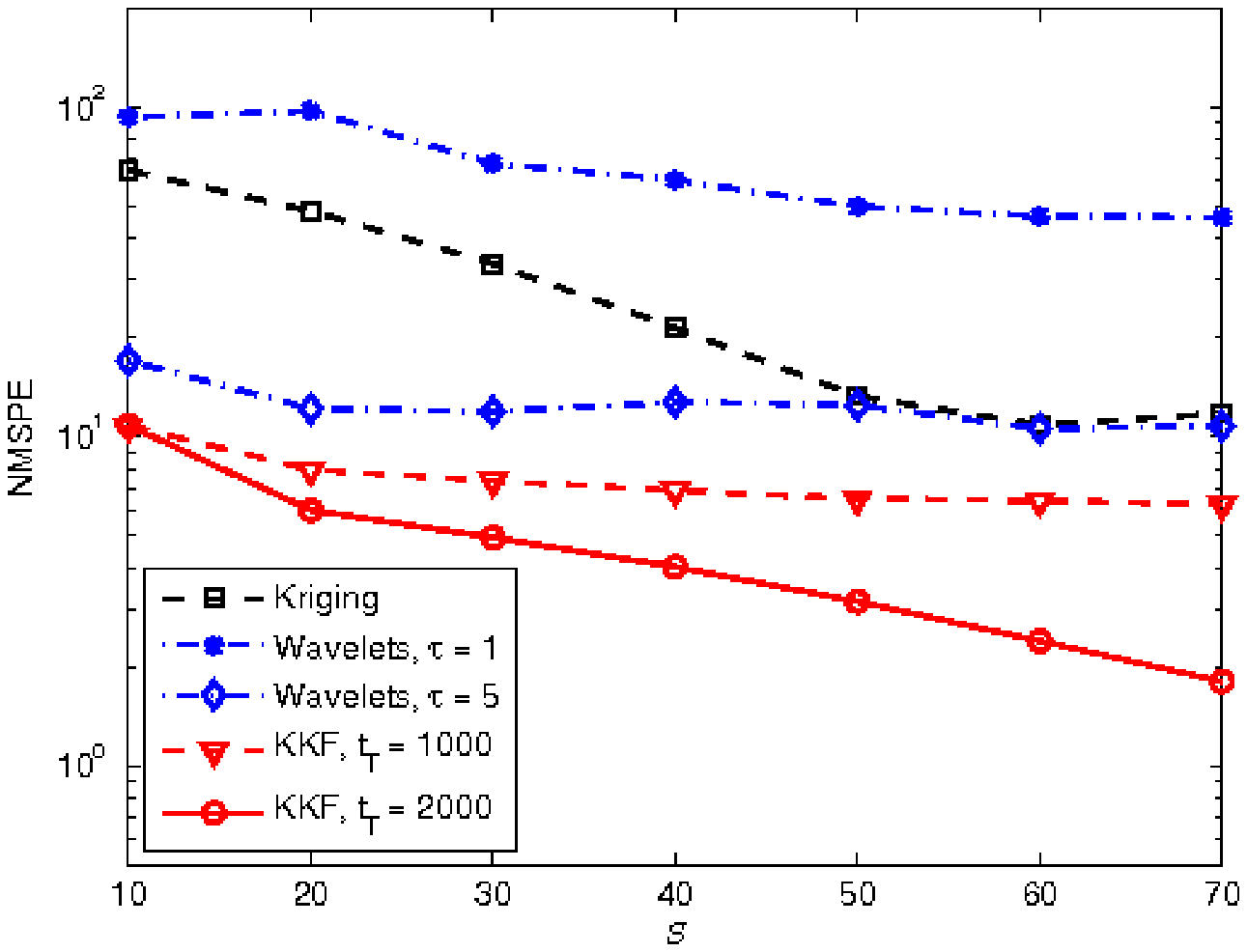}}
\medskip
\end{minipage}
\vspace{-0.4cm}
\caption{Delay cartography using the NZ-AMP dataset~\cite{amp}, which includes 
path delays collected over a month for an IP network where $P=186$ and 
$N=30$~\cite{rajassp12}. Normalized mean-square prediction error (NMSPE) 
as a function of $S$.
(Left) Random path selection; and (Right) ``Optimal'' path selection, 
that is, using heuristic or approximate algorithms specified for each algorithm. 
Observe further that the performance of the KKF
improves as the length of the training interval $t_T$ increases.}
\vspace{-0.8cm}
\label{fig:delay_perf} 
\end{figure}

The KKF framework for dynamic network delay cartography has several attractive features. 
First, the KKF yields the LMMSE estimate even for non-Gaussian distributed noise. 
The Kalman filter step also allows for a $\tau$-step 
prediction given by $\hat{\bd}_{t+\tau} = \hat{\bchi}_t$, 
which can be useful for preemptive routing and congestion control algorithms, as well 
as for extrapolating missing measurements. Second, the KKF framework provides a metric, 
namely the error covariance matrix $\bM^{\bar{s}}_t$, for choosing the paths to be 
measured at each $t$, which define the selection matrix $\St$. In the present setting, 
it turns out that the D-optimal design metric $\log \det \bM^{\bar{s}}_t$ is monotonic 
and supermodular with respect to the set $\cS$~\cite{rajassp12}. Thus, a simple greedy 
algorithm with complexity $\mathcal{O}(PS^3)$ can be employed to find the set of paths 
that are at least 63\% optimal \cite{nemh}; see Fig. \ref{fig:delay_perf}.
Consequently, the technique can be readily applied to large-scale networks 
since the complexity increases only linearly with $P$. 
The framework also admits related problem formulations such as selecting the best set of 
monitors (nodes) capable of measuring delay on all its outgoing paths. 
This represents a significant departure from state-of-the-art delay 
prediction/tracking methods~\cite{nk,coates}, where path selection 
is heuristic. Note that training is required to estimate the model 
parameters $\ceta$  and $\alpha$. To this end, empirical 
estimation techniques similar to those in~\cite{myers76} can be adapted 
to the present case. 

% % % % % % % % % % % % % % % % % % % % % % % % % % % % % % % % % % % % % % % %
%                         Section III                                         %
% % % % % % % % % % % % % % % % % % % % % % % % % % % % % % % % % % % % % % % %

\section{Dynamic Anomalography}
\label{sec:anomalography}

This section switches gears to \textit{anomalography}, the problem
of unveiling and mapping-out network traffic anomalies across flows and time
given link-level traffic measurements. This is a 
crucial monitoring task towards engineering network traffic, since 
anomalies can result in congestion and limit QoS provisioning. % of the end users.

% % % % % % % % % % % % % % % % % % % % % % % % % % % % % % % % % % % % % % % %
%                        Subsection III-A                                     %
% % % % % % % % % % % % % % % % % % % % % % % % % % % % % % % % % % % % % % % %

\subsection{Traffic modeling}
\label{ssec:model}

\begin{figure}[ht]
\centering
 \centerline{\epsfig{figure=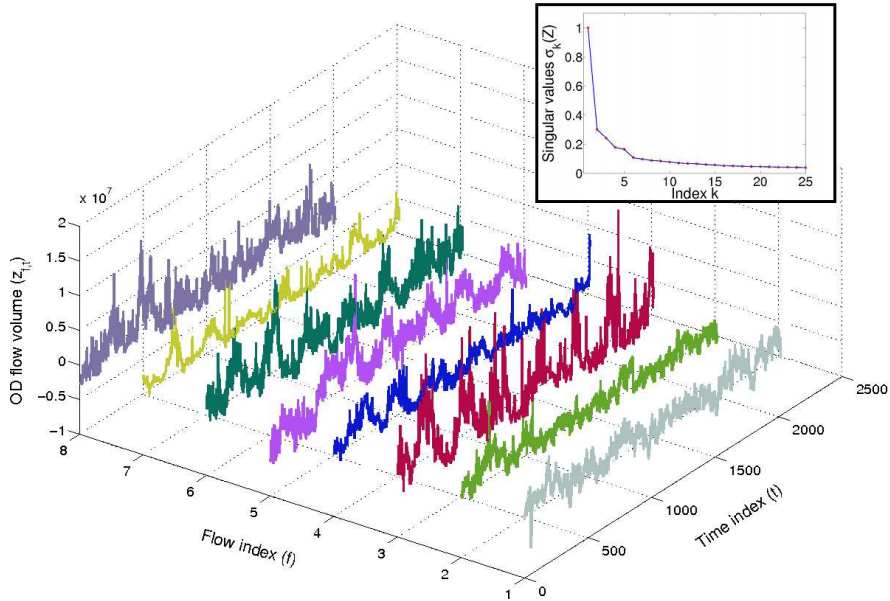,width=0.7\textwidth}}
\vspace{-0.4cm}\caption{Volumes of $6$ representative (out of $121$ total) OD flows, taken from
the operation of Internet-2 during a seven-day period~\cite{Internet2}. Temporal
periodicities and correlations across flows are apparent. As expected, in this case $\bZ$
can be well approximated by a low-rank matrix, since its normalized singular
values decay rapidly to zero.}
\vspace{-0.5cm} \label{fig:fig_flows}
\end{figure}

Consider a backbone IP network where $\cal{N}$ and $\mathcal{L}$ denote the sets of
nodes (routers) and physical links of cardinality $|\calN|=N$ and $|\mathcal{L}|=L$, respectively. 
The operational goal of the network is to transport a set of 
OD traffic flows $\cal{F}$ (with 
$|\mathcal{F}| = F$) associated with specific OD (ingress-egress router) pairs. 
Single-path routing is adopted here, meaning a given flow's traffic is carried through multiple
links connecting the corresponding source-destination pair along a single path. Accordingly, 
over a discrete time horizon $t \in [1,T]$ the measured link counts 
$\bX:=[x_{l,t}]\in\mathbb{R}^{L \times T}$ and (unobservable) OD flow traffic matrix
$\bZ:=[z_{f,t}]\in\mathbb{R}^{F \times T}$, are thus related through $\bX=\bR\bZ$ [cf. \eqref{eq:ltra}].
Unless otherwise stated, the routing matrix $\bR$ is assumed given, 
since it can be otherwise estimated using \textit{traceroute} or topology
inference algorithms~\cite{high-rank}. 
It is also fat, as for backbone networks the number
of OD flows is much larger than the number of physical links $(F\gg L)$.
A cardinal property of the traffic matrix is noteworthy. 
Common temporal patterns across OD traffic flows in addition to their almost 
periodic behavior, render most rows (respectively columns) of the traffic 
matrix linearly dependent, and thus $\bZ$
typically has \textit{low rank}. This intuitive property has been extensively validated
with real network data; see Fig. \ref{fig:fig_flows} and e.g.,~\cite{lakhina}.

It is not uncommon for some of the OD flow rates to experience unexpected abrupt changes. These
so-termed \textit{traffic volume anomalies} are typically due to
(unintentional) network equipment misconfiguration or outright failure, unforeseen behaviors following 
routing policy modifications, or,
cyberattacks (e.g., DoS attacks) which aim at compromising the
services offered by the network~\cite{zggr05,lakhina}.
Let $a_{f,t}$ denote the unknown amount of anomalous traffic in flow $f$ at
time $t$. Explicitly accounting for the presence of anomalous flows, the measured 
traffic carried by link $l$ is then given by
%
%\begin{equation}
$y_{l,t}=\sum_{f\in\cal{F}}r_{l,f}(z_{f,t}+a_{f,t}) + \epsilon_{l,t},~t=1,...,T$, %\label{eq:y_lt}
%\end{equation}
%
where the noise variables $\epsilon_{l,t}$ capture measurement errors
and unmodeled dynamics. Traffic volume anomalies are (unsigned) sudden
changes in OD flow's traffic, and as such their effect can span
multiple links in the network. A key difficulty in unveiling anomalies
from link-level measurements only is that  oftentimes, clearly
discernible anomalous spikes in the flow traffic can be masked through
``destructive interference'' of the superimposed OD flows~\cite{lakhina}. 
An additional challenge stems from missing link-level measurements $y_{l,t}$,
an unavoidable operational reality
affecting most traffic engineering tasks that rely on (indirect) measurement 
of traffic matrices~\cite{zrwq09,Roughan}. 
To model missing link measurements, collect the tuples $(l,t)$ associated with the
available observations $y_{l,t}$ in the set $\Omega \subseteq [1,2,...,L] \times
[1,2,...,T]$. Introducing the matrices $\bY:=[y_{l,t}],
\bE:=[\epsilon_{l,t}]\in\mathbb{R}^{L \times T}$, and $\bA:=[a_{f,t}]
\in\mathbb{R}^{F \times T}$, the (possibly incomplete) set of link-traffic measurements
can be expressed in compact matrix form as
\begin{equation}
\cP_{\Omega}(\bY)=\cP_{\Omega}(\bX + \bR\bA + \bE)\label{eq:Y}
\end{equation}
where the sampling operator $\cP_{\Omega}(.)$ sets
the entries of its matrix argument not in $\Omega$ to zero, and keeps the rest
unchanged. Since the objective here is not
to estimate the OD flow traffic matrix $\bZ$, \eqref{eq:Y} is expressed in
terms of the nominal (anomaly-free) link-level traffic rates $\bX$, which inherits
the low-rank property of $\bZ$. Anomalies in $\bA$ are expected 
to occur sporadically over time, and last for a short time
relative to the (possibly long) measurement interval $[1,T]$.
In addition, only a small fraction of the flows is supposed to be anomalous at a any given
time instant. This renders the anomaly traffic matrix $\bA$ 
\textit{sparse} across both rows (flows) and columns (time).

% % % % % % % % % % % % % % % % % % % % % % % % % % % % % % % % % % % % % % % %
%                        Subsection III-B                                     %
% % % % % % % % % % % % % % % % % % % % % % % % % % % % % % % % % % % % % % % %

\subsection{Unveiling anomalies via sparsity and low rank}
\label{ssec:unveil}

Given link-level traffic measurements $\cP_{\Omega}(\bY)$
adhering to~\eqref{eq:Y}, \textit{dynamic anomalography} is a critical network monitoring task 
that aims at accurately estimating the anomaly matrix $\bA$. As argued next, capitalizing
on the sparsity of $\bA$ and the low-rank property of $\bX$ will be
instrumental in achieving this ambitious goal. 
From a network cartography vantage point, the resultant estimated map 
$\hat{\bA}$ offers a depiction of the network's ``health state'' 
along both the flow and time dimensions. 
If $|\hat{a}_{f,t}|>0$, the $f$-th flow at time $t$ is deemed anomalous, otherwise it is healthy. 
This joint estimation-detection task not only allows one to identify
the time of the anomaly in addition to the affected flows, but also to estimate its magnitude
which hints to the importance of the anomaly event. 
By examining $\bR$ the network operator can immediately determine the links carrying
the anomalous flows. Subsequently, planned contingency measures involving traffic-engineering algorithms
can be implemented to address network congestion. 

The low-rank property of the traffic matrix $\bZ$ (and $\bX$) is at the heart of 
the seminal network anomaly detection approach in~\cite{lakhina}. In the absence of missing 
data, the method therein adopts principal component analysis (PCA) 
to decompose the link traffic $\bY=[\by_1,\ldots,\by_T]$ into 
nominal and anomalous components (also known as modeled and residual traffic). For instance, if most of
the variance in $\bY$ is captured by $r\ll \min(L,T)$ dominant principal components,
then by construction the nominal subspace $\mathcal{S}_n$ is spanned by 
the $r$ dominant right singular vectors
of $\bY'$ (cf. the low rank assumption). Naturally, the anomalous 
subspace $\mathcal{S}_a$ corresponds to the orthogonal complement, i.e., 
$\mathcal{S}_a:=\mathcal{S}_n^\perp$. In the operational phase, an anomaly is declared at time $t$
when $\|\mathbf{P}_{\mathcal{S}_a}\by_t\|_2^2$ exceeds a given threshold, where 
$\mathbf{P}_{\mathcal{S}_a}$ is an orthogonal projection matrix onto $\mathcal{S}_a$. 
Subsequently, a single anomalous flow is identified after running a greedy algorithm,
and an estimate of the amount of anomalous traffic is obtained as a byproduct.
Likewise, the \textit{spatial} approach within the \textit{network anomography} framework~\cite{zggr05}
forms the matrix $\mathbf{P}_{\mathcal{S}_a}\bY$ of link anomalies, thus exploiting the 
correlation between traffic across different links. \textit{Temporal} approaches obtain 
link anomalies as $\bY\bT$ instead, where $\bT$ is a linear operator which 
judiciously filters the traffic time series per link (implementing
an ``anomaly-pass'' filter). Several choices for $\bT$ are proposed to this end, 
based on different forms of temporal analysis including autoregressive integrated
moving average (ARIMA), wavelets, and fast Fourier transform (FFT). Different from~\cite{lakhina},
the inference algorithm in~\cite{zggr05} capitalizes on the sparsity of $\bA$ 
to estimate the anomaly map by e.g., solving in the spatial case
\begin{equation*}
\hat{\bA}:=\arg\min_{\bA}\|\bA\|_1,\quad \textrm{s. t. }\: 
\mathbf{P}_{\mathcal{S}_a}\bY=\bR\bA.
\end{equation*}
Network anomography algorithms can be extended to accommodate routing changes across 
time; see~\cite{zggr05} for further details and comprehensive performance tests. 

Recently, a natural estimator leveraging the low rank property of $\bX$ and 
the sparsity of $\bA$ was put forth in~\cite{tsp_rankminimization_2012}, which can
be found at the crossroads of CS~\cite{Do06CS} and timely low-rank plus
sparse matrix decompositions~\cite{CLMW09,CSPW11}. The idea is to fit 
the incomplete data $\cP_{\Omega}(\bY)$ to the model $\bX + \bR \bA$ [cf. \eqref{eq:Y}] in 
the LS error sense, as well as minimize the
rank of $\bX$, and the number of nonzero entries of $\bA$ measured by its $\ell_0$-(pseudo)
norm. Unfortunately, albeit natural both rank and $\ell_0$-norm criteria
are in general NP-hard to optimize.
Typically, the nuclear norm $\|\bX\|_*:=\sum_{k}\sigma_k(\bX)$ ($\sigma_k(\bX)$
denotes the $k$-th singular value of $\bX$) and the $\ell_1$-norm $\|\bA\|_1$
are adopted as surrogates~\cite{F02,CT05}, since they are the closest \textit{convex}
approximants to $\textrm{rank}(\bX)$ and $\|\bA\|_0$, respectively.
Accordingly, one solves
\begin{equation}
\min_{\{\bX,\bA\}} \|\cP_{\Omega}(\bY - \bX -
\bR\bA)\|_{F}^{2} +\lambda_{\ast}\|\bX\|_{*} + \lambda_1\|\bA\|_1\label{eq:p1}
\end{equation}
where $\lambda_*,\lambda_1\geq 0$ are rank- and sparsity-controlling parameters. 
%When an estimate $\hat{\sigma}_v^2$ of the noise variance is available,
%guidelines for selecting $\lambda_*$ and $\lambda_1$ have been proposed in~\cite{zlwcm10}.
While a non-smooth optimization problem, being convex \eqref{eq:p1} is appealing. 
An efficient accelerated proximal gradient algorithm with
quantifiable iteration complexity was developed to unveil network 
anomalies~\cite{tit_recovery_2012}. Interestingly, \eqref{eq:p1}
also offers a cleansed estimate of the link-level traffic $\hat{\bX}$, that
could be subsequently utilized for network tomography tasks.
In addition, \eqref{eq:p1} \textit{jointly} exploits the spatio-temporal
correlations in the link traffic as well as the sparsity of the anomalies, 
through an optimal single-shot estimation-detection procedure that has been shown to outperform 
the algorithms  in~\cite{lakhina} and~\cite{zggr05} (that decouple the estimation and
detection steps).

\begin{figure}[ht]
\begin{minipage}[b]{0.48\linewidth}
  \centering
  \centerline{\includegraphics[width=0.9\linewidth, height=2 in]{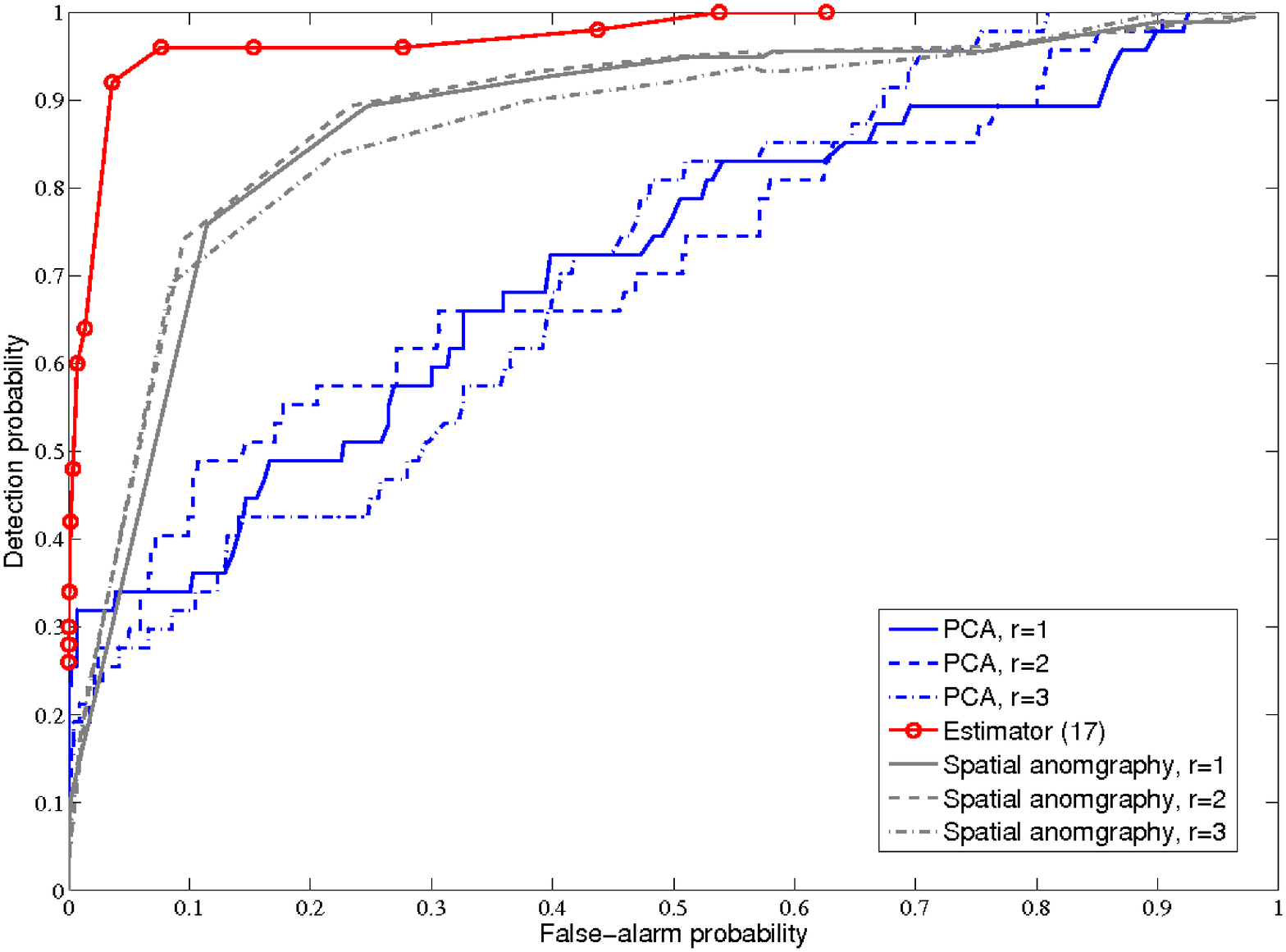}}
\medskip
\end{minipage}
\hfill
\begin{minipage}[b]{.48\linewidth}
  \centering
  \centerline{\includegraphics[width=0.9\linewidth, height=2 in]{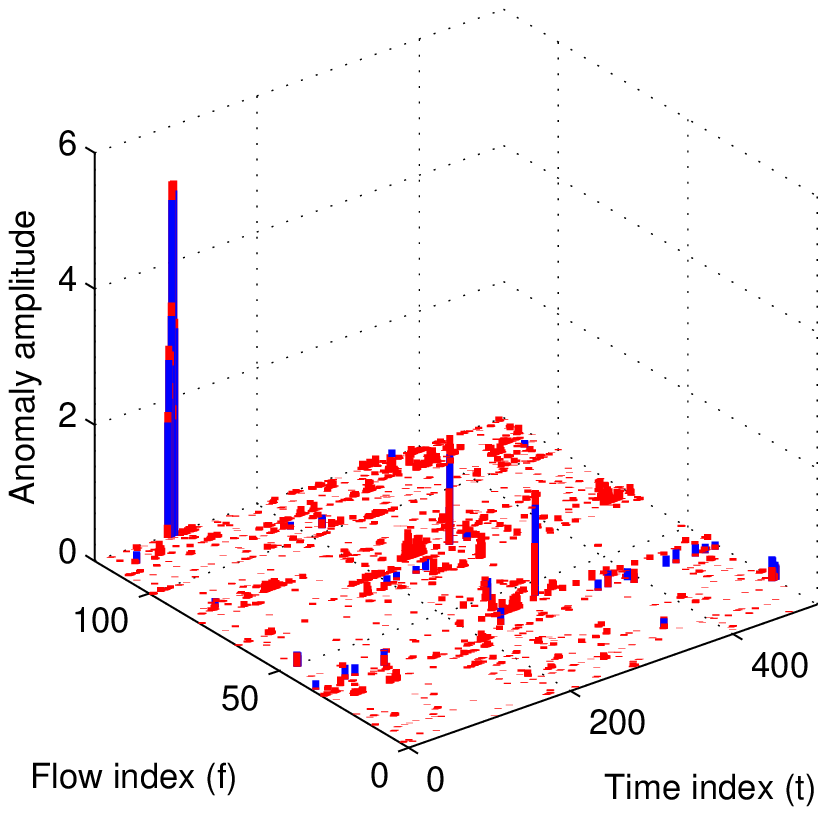}}
\medskip
\end{minipage}
\vspace{-0.4cm}
\caption{Unveiling anomalies from Internet-2 data~\cite{Internet2}. (Left) ROC curve comparison
between \eqref{eq:p1} and the PCA methods in~\cite{lakhina,zggr05}, for different values of
$r:=\textrm{dim}(\mathcal{S}_n)$. Leveraging sparsity and low rank jointly leads
to improved performance. (Right) In red, the estimated anomaly map $\hat{\bA}$ 
obtained via \eqref{eq:p1} superimposed to the ``true'' 
anomalies shown in blue~\cite{dyn_anomal}.}
\vspace{-0.4cm}
\label{fig:anomographt} 
\end{figure}

Before moving on to distributed implementations, it is instructive
to elaborate on the generality of \eqref{eq:p1}.
When there is no missing data and $\bX=\mathbf{0}_{L\times T}$, one is
left with an under-determined sparse signal recovery problem
typically encountered with CS; see e.g.,~\cite{Do06CS}. The
decomposition $\bY=\bX+\bA$ corresponds to principal component
pursuit (PCP), also referred to as robust PCA~\cite{CLMW09,CSPW11}. For the idealized
noise-free setting ($\bE=\mathbf{0}_{L\times T}$), sufficient conditions for exact recovery
of the unknowns are
available for both of the aforementioned special cases~\cite{CT05,CLMW09,CSPW11}.
However, the superposition of a low-rank plus a \textit{compressed}
sparse matrix in \eqref{eq:Y} further challenges identifiability
of $\{\bX,\bA\}$; see~\cite{tit_recovery_2012} for early results. 
Going back to the CS paradigm, even when $\bX$ is nonzero one could envision a variant
where the measurements are corrupted with correlated (low-rank)
noise~\cite{Vaswani_Allerton_11}. Last but not least, when
$\bA=\mathbf{0}_{F\times T}$ and $\bY$ is noisy, the recovery of
$\bX$ subject to a rank constraint is nothing but PCA --
arguably, the workhorse of high-dimensional data
analytics. This same formulation is adopted for low-rank
matrix completion, to impute the missing entries of a low-rank
matrix observed in noise, i.e., $\cP_{\Omega}(\bY) =
\cP_{\Omega}(\bX +\bE)$~\cite{candes_moisy_mc}.

% % % % % % % % % % % % % % % % % % % % % % % % % % % % % % % % % % % % % % % %
%                        Subsection III-C                                     %
% % % % % % % % % % % % % % % % % % % % % % % % % % % % % % % % % % % % % % % %

\subsection{In-network distributed processing}
\label{ssec:inp}

Implementing \eqref{eq:p1} presumes that network nodes continuously
communicate their link traffic measurements to a central monitoring station, 
which uses their aggregation in 
$\cP_{\Omega}(\bY)$ to unveil anomalies.
While for the most part this is the prevailing operational paradigm adopted
in current networks, it is fair to say 
there are limitations associated with this architecture.
For instance, fusing all this information may entail
excessive protocol overheads. Moreover, minimizing the exchanges of raw measurements
may be desirable to reduce unavoidable communication errors that translate to missing data. 
Solving \eqref{eq:p1} centrally raises robustness concerns as well,
since the central monitoring station represents an isolated point of failure. 

These reasons motivate well devising \emph{fully-distributed} iterative algorithms
for dynamic anomalography, embedding the network
anomaly detection functionality to the routers. In a nutshell, per iteration 
nodes $n\in\cN$ carry out simple computational tasks locally, 
relying on their own link count measurements (a submatrix $\bY_{n}$
within $\bY = [\bY_1',\ldots,\bY_N']'$ corresponding to router $n$'s links). 
Subsequently, local estimates are refined after exchanging messages 
only with directly connected neighbors, which facilitates percolation
of local information to the whole network. The end goal is 
for network nodes to consent on a global map of network anomalies $\hat{\bA}$, 
and attain (or at least come close to) the estimation performance
of the centralized counterpart \eqref{eq:p1} which has all data $\cP_{\Omega}(\bY)$
available.

Problem \eqref{eq:p1} is not amenable for distributed implementation
due to the non-separable nuclear norm present in the cost
function. If an upper bound $\textrm{rank}(\hat\bX)\leq
\rho$ is a priori available [recall $\hat\bX$ is the estimated link-level traffic 
obtained via \eqref{eq:p1}], \eqref{eq:p1}'s search space is
effectively reduced and one can factorize the decision variable as
$\bX=\bP\bQ'$, where $\bP$ and $\bQ$ are $L \times \rho$ and $T
\times \rho$ matrices, respectively. Again, it is possible to interpret 
the columns of $\bX$ (viewed as points in $\mathbb{R}^L$) as belonging to
a low-rank nominal subspace $\mathcal{S}_n$, spanned by the columns of $\bP$. The rows of $\bQ$
are thus the projections of the columns of $\bX$ onto $\mathcal{S}_n$.
Next, consider the following alternative characterization of
the nuclear norm (see e.g.~\cite{Recht_Parallel_2011})
\begin{equation}\label{eq:nuc_nrom_def}
\|\bX\|_*:=\min_{\{\bP,\bQ\}}~~~ \frac{1}{2}\left(\|\bP\|_F^2+\|\bQ\|_F^2 \right),\quad
\text{s. t.}~~~ \bX=\bP\bQ'
\end{equation}
where the optimization is over all possible bilinear factorizations
of $\bX$, so that the number of columns $\rho$ of $\bP$ and
$\mathbf{Q}$ is also a variable. Leveraging \eqref{eq:nuc_nrom_def}, 
the following reformulation of \eqref{eq:p1} provides an important first step 
towards obtaining a distributed anomalography algorithm
\begin{equation}
\min_{\{\bP,\bQ,\bA\}}\sum_{n=1}^N \left[ \|\cP_{\Omega_n}(\bY_n 
-\bP_n\bQ'-\bR_n\bA)\|_F^2+\frac{\lambda_{*}}{2N}\left(N\|\bP_n\|_F^2 + \|\bQ\|_F^2 \right) 
+ \frac{\lambda_1}{N}\|\bA\|_1\right]\label{eq:p2}
\end{equation}
which is non-convex due to the bilinear terms $\bP_n\bQ'$, and where
$\bR :=\left[\bR_1',\ldots,\bR_N'\right]^\prime$ is partitioned into local
routing tables available per router $n$.
Adopting the separable Frobenius-norm 
regularization in \eqref{eq:p2} comes with no loss of optimality relative to \eqref{eq:p1}, 
provided $\textrm{rank}(\hat\bX)\leq\rho$. By finding the global minimum of \eqref{eq:p2}
[which could have considerably less variables than \eqref{eq:p1}], 
one can recover the optimal solution of \eqref{eq:p1}. But since \eqref{eq:p2} is
non-convex, it may have stationary points which need not be globally
optimum. As asserted in~\cite[Prop. 1]{tsp_rankminimization_2012} however, if a stationary point
$\{\bar{\bP},\bar{\bQ},\bar{\bA}\}$ of \eqref{eq:p2} satisfies
$\|\cP_{\Omega}(\bY - \bar{\bP}\bar{\bQ}' - \bar{\bA})\| < \lambda_*$, then
$\{\hat\bX:=\bar{\bP}\bar{\bQ}',\hat{\bA}:=\bar{\bA}\}$ is the globally optimal solution of
\eqref{eq:p1}.  Note that for sufficiently small $\rho$ the residual
$\|\cP_{\Omega}(\bY - \bar{\bP}\bar{\bQ}' - \bar{\bA})\|$
becomes large, and the qualification inequality is violated [unless $\lambda_*$ is large
enough, in which case a sufficiently low-rank solution to \eqref{eq:p1} is expected]. The condition
on the residual implicitly enforces $\textrm{rank}(\hat\bX) \leq \rho$, which is necessary for
the equivalence between \eqref{eq:p1} and \eqref{eq:p2}.

To decompose the cost in \eqref{eq:p2}, in which summands inside the square brackets 
are coupled through the global variables $\{\bQ,\bA\}$, 
introduce auxiliary copies $\{\bQ_n,\bA_n\}_{n=1}^N$
representing local estimates of $\{\bQ,\bA\}$, one per node $n$. These
local copies along with \textit{consensus} constraints yield the distributed estimator
\begin{align}
\min_{\{\bP_n,\bQ_n,\bA_n\}}& \sum_{n=1}^N \left[\|\cP_{\Omega_n}(\bY_n 
-\bP_n\bQ_n'-\bR_n\bA_n)\|_F^2+ 
\frac{\lambda_{*}}{2N}\left(N\|\bP_n\|_F^2 + \|\bQ_n\|_F^2 \right) +
\frac{\lambda_1}{N}\|\bA_n\|_1\right] \label{eq:p3}\\
\text{s. t.} &\quad \bQ_n=\bQ_m,\:\bA_n=\bA_m\quad m\textrm{ linked with } n\in\mathcal{N}\nonumber
\end{align}
which is equivalent to \eqref{eq:p2} provided the 
network topology graph is connected. Even though consensus 
is a fortiori imposed within neighborhoods, it extends to
the whole (connected) network and local estimates agree on the
global solution of \eqref{eq:p2}. Exploiting the separable structure
of \eqref{eq:p3}, a general framework for in-network sparsity-regularized
rank minimization was put forth in~\cite{tsp_rankminimization_2012}. Specifically,
distributed iterations were obtained after adopting the alternating-direction method of 
multipliers (ADMM), an iterative Lagrangian method well-suited for
parallel processing~\cite{Bertsekas_Book_Distr}. In a nutshell, 
local tasks per iteration $k=1,2,\ldots$ entail
solving small unconstrained quadratic programs to refine the normal subspace $\bP_n[k]$, 
in addition to soft-thresholding operations to update the anomaly maps $\bA_n[k]$ per router.
Each iteration, routers exchange their estimates $\{\bQ_n[k],\bA_n[k]\}$ only with directly
connected neighbors. This way the communication overhead remains affordable, and independent of 
the network size $N$.

When employed to solve non-convex problems such as \eqref{eq:p3}, so far ADMM
offers no convergence guarantees. However, there is ample
experimental evidence in the literature that supports empirical
convergence of ADMM, especially when the non-convex problem at hand exhibits
``favorable'' structure. For instance, \eqref{eq:p3} is a
linearly constrained bi-convex problem with potentially good convergence properties 
-- extensive numerical tests in~\cite{tsp_rankminimization_2012} demonstrate that this is
indeed the case. While establishing convergence remains an open problem, one can still
prove that upon convergence the distributed iterations attain 
consensus and global optimality, offering the desirable centralized performance 
guarantees~\cite{tsp_rankminimization_2012}. 
%A noteworthy distributed PCA algorithm was developed in~\cite{hero_DPCA}, 
%and successfully applied to diagnose network anomalies as in~\cite{lakhina}.

% % % % % % % % % % % % % % % % % % % % % % % % % % % % % % % % % % % % % % % %
%                        Subsection III-D                                     %
% % % % % % % % % % % % % % % % % % % % % % % % % % % % % % % % % % % % % % % %

\subsection{Real-time anomaly trackers}
\label{ssec:online}

Monitoring of large-scale IP networks necessitates massive recollection 
of data which far outweigh the ability of modern computers to store and analyze them 
in real time. In addition, nonstationarities due to routing changes and missing data
further challenge identification of anomalies. In dynamic networks routing tables
are constantly readjusted to effect traffic load balancing and avoid
congestion caused by e.g., traffic anomalies. 
To account for slowly time-varing routing tables, let $\bR_t \in
\mathbb{R}^{L \times F}$ denote the routing matrix at time $t$. 
In this dynamic setting, the partially observed 
link counts at time $t$ adhere to
$\cP_{\Omega_t}(\by_t)=\cP_{\Omega_t}(\bx_t + \bR_t\ba_t +\bm{\epsilon}_t),~t=1,2,\ldots$, 
where the link-level traffic $\bx_t:=\bR_t\bz_t$. In general, routing
changes may alter a link load considerably by e.g.,
routing traffic completely away from a specific link. Therefore, even
though the OD flow vectors $\{\bz_t\}$ live in a low-dimensional
subspace, the same may not be true for the $\{\bx_t\}$
when the routing updates are major and frequent.
In backbone networks however, routing changes are sporadic relative
to the time-scale of data acquisition used for network monitoring tasks. 
For example, data collected from the operation of Internet-2 
network  reveals that only a few rows of 
$\bR_t$ change per week~\cite{Internet2}. It is thus safe to assume that
$\{\bx_t\}$ still lies in a low-dimensional subspace, and exploit the spatio-temporal correlations 
of the observations to identify the anomalies in real-time.

\begin{figure}[ht]
\begin{minipage}[b]{0.48\linewidth}
  \centering
  \centerline{\includegraphics[width=\linewidth, height=2 in]{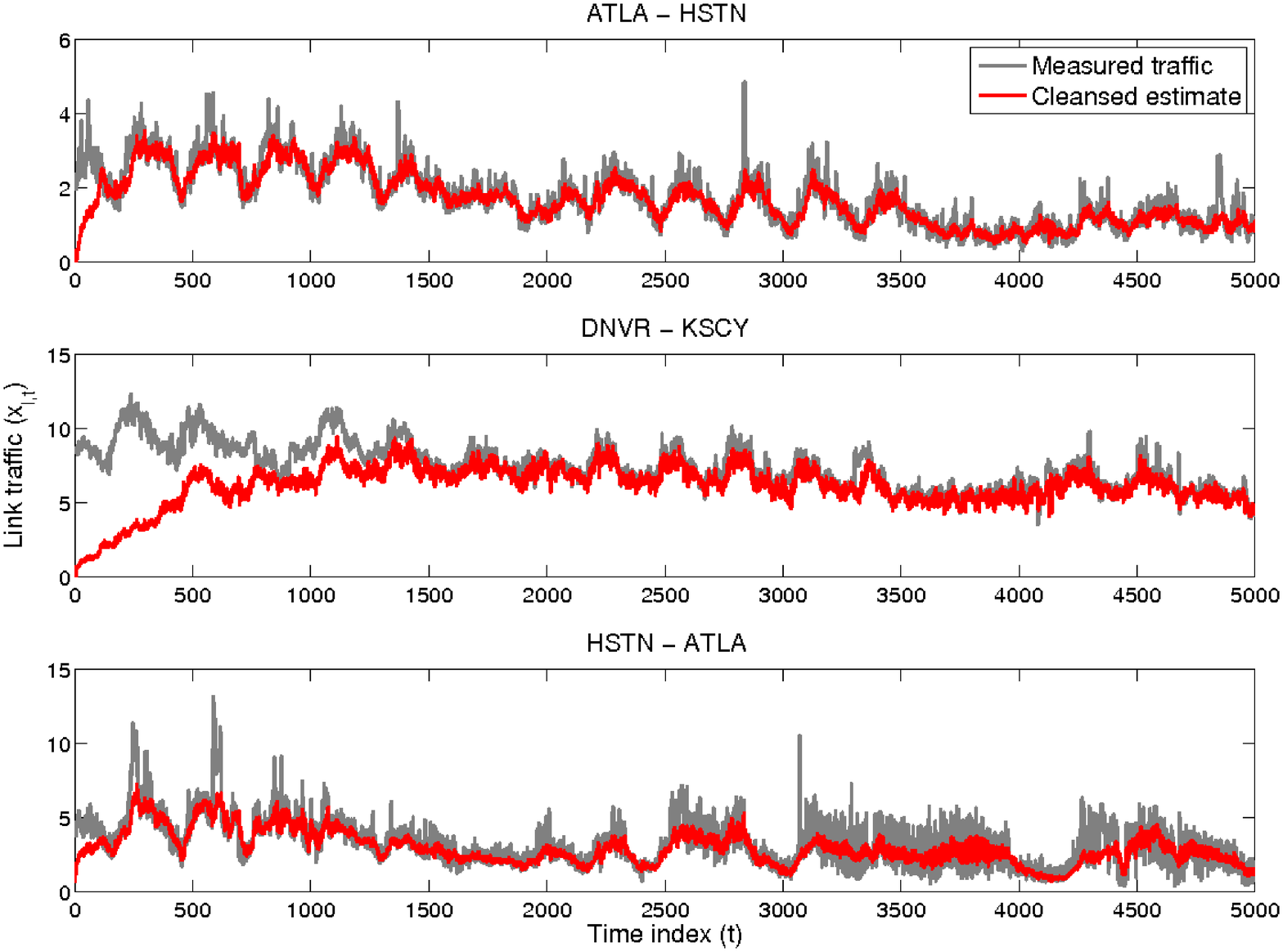}}
\medskip
\end{minipage}
\hfill
\begin{minipage}[b]{.48\linewidth}
  \centering
  \centerline{\includegraphics[width=\linewidth, height=2 in]{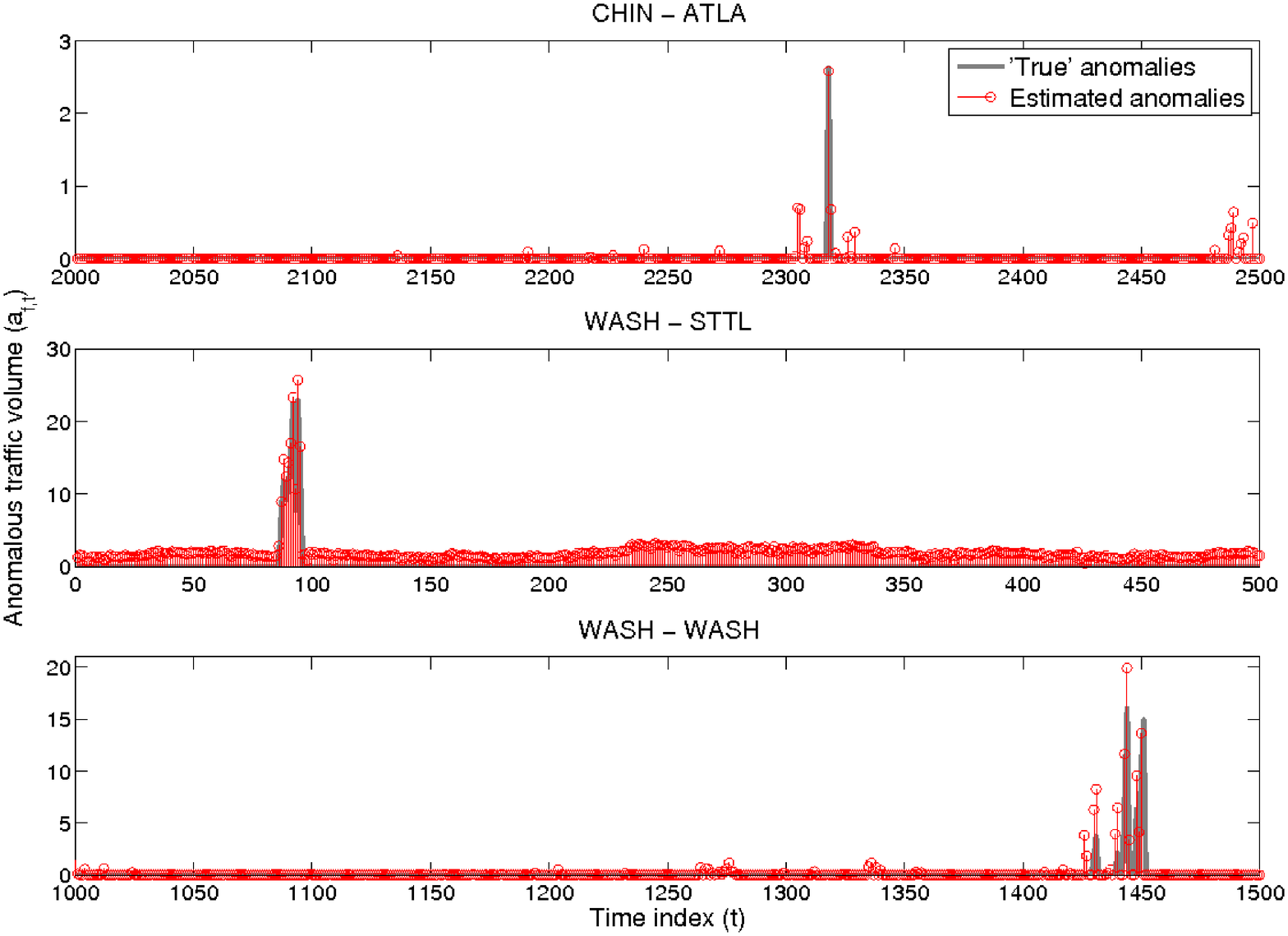}}
\medskip
\end{minipage}
\vspace{-0.4cm}
\caption{Unveiling anomalies in real time from Internet-2 data~\cite{Internet2}. (Left) Measured
link traffic and cleansed estimates for three representative links; and (Right) three rows
of the estimated anomaly map $\hat{\bA}$ corresponding to three anomalous flows~\cite{dyn_anomal}.}
\vspace{-0.4cm}
\label{fig:tracking} 
\end{figure}

On top of the previous arguments, in practice link measurements are acquired sequentially in
time, which motivates updating previously obtained estimates rather than
re-computing new ones from scratch each time a new datum becomes
available. The goal is then to recursively estimate $\{\hat{\bx}_t,\hat{\ba}_t\}$ at time
$t$ from historical observations $\{\cP_{\Omega_\tau}(\by_\tau)\}_{\tau=1}^t$,
naturally placing more importance on recent measurements. To this end, one
possible adaptive counterpart to~\eqref{eq:p2} is the exponentially-weighted 
LS estimator found by minimizing the empirical cost~\cite{dyn_anomal}
\begin{align}
\min_{\{\bP,\bQ,\bA\}} \sum_{\tau=1}^t \beta^{t-\tau}\left[ 
\|\cP_{\Omega_\tau}(\by_\tau-\bP\bq_\tau-\bR_\tau\ba_\tau) \|_2^2 + \frac{\lambda_{\ast}}{2
\sum_{u=1}^t \beta^{t-u}} \|\bP\|_F^2 +  \frac{\lambda_{\ast}}{2} \|\bq_\tau\|_2^2
+ \lambda_1 \|\ba_\tau\|_1 \right] \label{eq:adaptive_v1_est_lq}
\end{align}
in which $ 0< \beta \leq 1$ is the so-termed forgetting factor. When $\beta<1$ 
data in the distant past are 
exponentially downweighted,  which facilitates tracking network 
anomalies in nonstationary environments. For static
routing ($\bR_t=\bR$) and infinite memory $(\beta=1)$, 
the formulation~\eqref{eq:adaptive_v1_est_lq} coincides
with the batch estimator \eqref{eq:p2}. A provably convergent online algorithm for dynamic 
anomalography is developed in~\cite{dyn_anomal}, based on alternating minimization
of \eqref{eq:adaptive_v1_est_lq}; see Fig. \ref{fig:tracking}.
Each time a new datum is acquired, anomaly estimates are formed via the Lasso~\cite{Ti1996lasso}, 
and the low-rank nominal traffic subspace is refined using
recursive LS. For situations were reducing computational complexity is critical, 
an online stochastic gradient algorithm based on Nesterov's acceleration technique is developed
as well~\cite{dyn_anomal}.

Algorithms in~\cite{dyn_anomal} are closely related to timely robust
subspace trackers, which aim at estimating a low-rank subspace $\bP$
from grossly corrupted and possibly incomplete data, namely $\cP_{\Omega_t}(\by_t)=\cP_{\Omega_t}
(\bP\bq_t + \ba_t +\bm{\epsilon}_t),~t=1,2,\ldots$. In the absence of sparse ``outliers'' $\{\ba_t\}_{t=1}^\infty$,
an online algorithm based on incremental gradient descent on the Grassmannian manifold of subspaces
was put forth in~\cite{onlinetracking_bolzano10}. The second-order 
RLS-type algorithm in~\cite{petrels_chi12} extends the seminal projection
approximation subspace tracking (PAST) algorithm to handle missing data.
When outliers are present, robust counterparts can be found 
in~\cite{Vaswani_Allerton_11,increm_grad_grasmannian_bolzano12}.
Relative to all aforementioned works, the estimation problem \eqref{eq:adaptive_v1_est_lq} 
is more challenging due to the presence of the (compression) 
routing matrix $\bR_t$; see~\cite{tit_recovery_2012} for fundamental
identifiability issues related to the model~\eqref{eq:Y}.

% % % % % % % % % % % % % % % % % % % % % % % % % % % % % % % % % % % % % % %
%                         Section IV                                        %
% % % % % % % % % % % % % % % % % % % % % % % % % % % % % % % % % % % % % % %

\section{Broadening the network atlas}
\label{sec:atlas}

Additional cartography instances are outlined in this section, 
including anomalography from flow measurements
and network distance prediction.
To exemplify the development of sensing infrastructure
for situational awareness at the physical layer of wireless CR networks, 
the notion of RF cartography is introduced as well. All these problems can be tackled
through SP methods subsumed by \eqref{eq:p1}, namely PCP~\cite{CSPW11}, %~\cite{CLMW09,CSPW11}, 
low-rank matrix completion~\cite{candes_moisy_mc}, %~\cite{F02,candes_moisy_mc},
the Lasso~\cite{Ti1996lasso}, and non-parametric versions of basis pursuit~\cite{rf_splines}.

% % % % % % % % % % % % % % % % % % % % % % % % % % % % % % % % % % % % % % % %
%                        Subsection IV-B                                      %
% % % % % % % % % % % % % % % % % % % % % % % % % % % % % % % % % % % % % % % %

\vspace*{-0.3cm}
\subsection{Unveiling anomalies from flow data}
\label{ssec:netflow}

Since some networks nowadays collect OD flow (not link-level) measurements $z_{f,t}+a_{f,t}$
for at least part of their network (using e.g., the Netflow protocol), 
anomalies can be detected
using temporal decomposition and standard change-detection approaches per flow.
Leveraging the low-rank property of the traffic matrix and the sparsity
of anomalies, anomalography from OD flow measurements was formulated as 
the PCP matrix decomposition problem and
solved centrally in~\cite{ajwak10}; see also~\cite{tsp_rankminimization_2012} for a distributed
implementation of the PCP estimator aimed at scalable monitoring of networks.

% % % % % % % % % % % % % % % % % % % % % % % % % % % % % % % % % % % % % % % %
%                        Subsection IV-B                                      %
% % % % % % % % % % % % % % % % % % % % % % % % % % % % % % % % % % % % % % % %

\vspace*{-0.3cm}
\subsection{Network distance prediction}
\label{ssec:distance}

End-to-end \textit{network distance} information is critical towards enhancing QoS
in Internet applications such as content distribution and
peer-to-peer file sharing systems. Clients naturally prefer to establish
connections with ``closer'' network resources or servers that are likely
to respond faster. There are different metrics to quantify the distance
between a pair of network nodes. The most common choices are defined
in terms of latency (one-way delay and the so-termed round-trip time)
or router hop-counts. Unfortunately, either probing or passively
measuring all pairwise distances becomes infeasible in large-scale networks.
%since the number of desired measurements grows quadratically with the 
%number of nodes. 
Given those few affordable distance measurements, 
the problem of network distance prediction is to impute (that is interpolate) the missing entries 
in a highly-incomplete matrix of end-to-end distances.

If one collects the end-to-end latencies $d_{i,j}$
of source-sink pairs $(i,j)$ in a delay matrix $\bD:=[d_{i,j}] \in \mathbb{R}^{N \times N}$,
strong dependencies among path delays render $\bD$
low rank; see e.g.,~\cite{latencyprediction}
for an experimental validation with multiple datasets.
Intuitively, correlations among rows and columns of $\bD$  emerge because nearby nodes (e.g., those
belonging to a common subnetwork) are connected to every other node through
paths with significant overlap, possibly sharing common bottleneck links. 
The low-rank property of $\bD$  along with the
distributed-processing requirements of large-scale networks, motivated
decentralized matrix-factorization~\cite{latencyprediction} and
nuclear-norm minimization~\cite{tsp_rankminimization_2012} algorithms
for network distance prediction. Different from schemes 
based on Euclidean embedding via multi-dimensional
scaling~\cite{vivaldi}, low-rank modeling does not require distances in $\bD$ 
to be symmetric and satisfy the triangle inequality -- properties that are 
oftentimes violated by network-related distances~\cite{tiv}.

To avoid the excessive overhead of active probing mechanisms, one can leverage network
monitors that passively observe router hop-counts from traffic
traversing those monitored links; see e.g.,~\cite{high-rank} and references
therein. Collect these hop-count measurements
in the matrix $\bH:=[h_{m,n}] \in \mathbb{N}^{M \times N}$,
where $M$ is the number of monitors, and $N$ ($\gg M$) the total hosts observed.
Because monitor $m$ only observes a fraction of the total network
traffic, $\bH$ will be depleted with missing entries.
Despite typically having $\textrm{rank}(\bH)=M$, $\bH$ consists of
low-rank column blocks, each corresponding to a subnetwork with access to the Internet
core through a single border router. Recognizing this structure, a \emph{high-rank} 
matrix completion algorithm that performs
subspace clustering of incomplete hop-count data was put forth in~\cite{high-rank}, 
and shown to attain good performance both in theory and practice.

Different from the dynamic network delay cartography problem considered in 
Sec. \ref{ssec:dnk}, network distance prediction approaches do not account for the temporal 
variations in the delays, and typically rely on batch imputation of the
distance matrix of interest. The techniques used in Sec.~\ref{ssec:dnk} do not apply in this 
context either, since some path delays are never observed, and thus it is 
impossible to estimate the spatial covariance matrices (such as $\ceta$ 
and $\cnu$) completely.

% % % % % % % % % % % % % % % % % % % % % % % % % % % % % % % % % % % % % % % %
%                        Subsection IV-C                                      %
% % % % % % % % % % % % % % % % % % % % % % % % % % % % % % % % % % % % % % % %

\vspace*{-0.3cm}
\subsection{RF cartography}
\label{ssec:rf}

In the domain of spectrum sensing for CR networks, 
RF cartography amounts to constructing in a distributed fashion: m1) global
power spectral density (PSD) maps capturing the distribution of radiated power across space,
time, and frequency; and m2) local channel gain (CG) maps offering the propagation 
medium per frequency from each node to any point in space. These maps enable identification of
opportunistically available spectrum bands for re-use and handoff operation;
as well as localization, transmit-power estimation, and tracking of 
primary user activities. While the focus here is on the
construction of PSD maps, the interested reader is referred to~\cite{tut_rf_cartog}
for a tutorial treatment on CG cartography.

A cooperative approach to RF cartography was introduced in
\cite{bazerque}, that builds on a basis
expansion model of the PSD map $\Phi(\bbx,f)$ across space $\bbx\in\mathbb{R}^2$, and frequency $f$. 
Spatially-distributed CRs collect
smoothed periodogram samples of the received signal at given sampling frequencies,
based on which they want to determine the unknown expansion coefficients. Introducing
a virtual spatial grid of candidate source locations, the estimation task
can be cast as a linear LS problem with an augmented vector of unknown parameters.
Still, the problem complexity (or effective degrees of freedom) can be controlled by capitalizing 
on two forms of sparsity: the first one introduced by the
narrow-band nature of transmit-PSDs relative to the broad swaths
of usable spectrum; and the second one emerging from sparsely located
active radios in the operational space (due to the grid artifact). Nonzero
entries in the parameter vector sought correspond to spatial location-frequency band pairs
corresponding to active transmissions.
All in all, estimating the PSD map and locating the active transmitters as a byproduct
boils down to a variable selection problem. This motivates
well employment of the Lasso for distributed sparse linear regression~\cite{tsp_rankminimization_2012}, an
estimator also subsumed by \eqref{eq:p1} 
when $\bX=\mathbf{0}_{L\times T}$, $T=1$, and the regression matrix
$\bR$ has a specific structure that depends on the chosen bases and path-loss propagation
model. 

\begin{figure}[ht]
\centering
  \centerline{\epsfig{figure=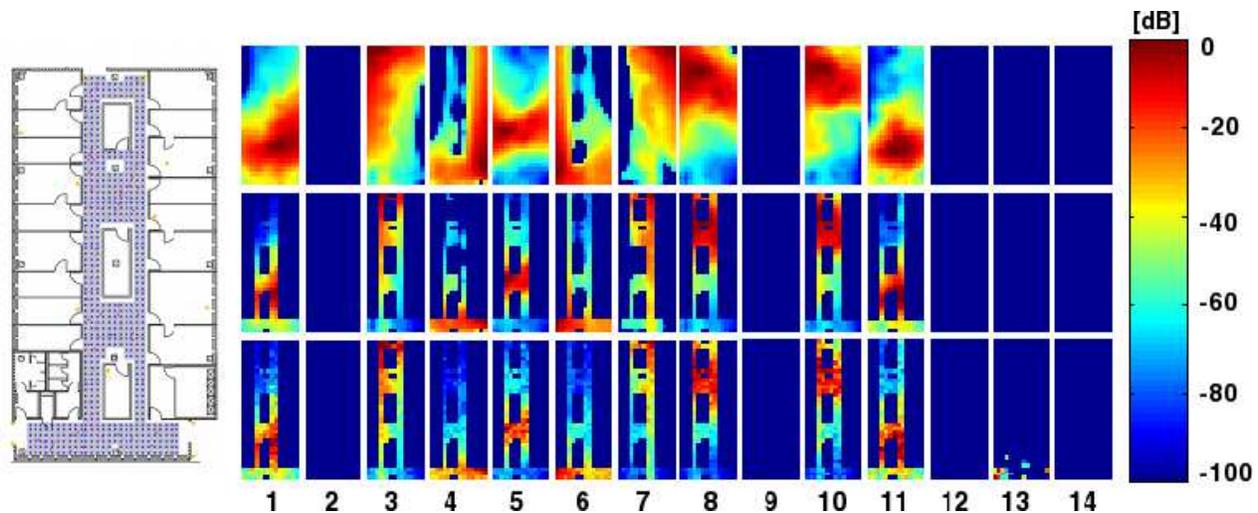,width=\linewidth}}
\caption{Spline-based RF cartography using the dataset~\cite{crowdad}. 
(Left) Detailed floor plan schematic including the location
of $N=166$ sensing radios; (Right-bottom)
original measurements spanning $14$ frequency sub-bands; (Right-center) estimated maps over the
surveyed area; and (Right-top) extrapolated maps. The proposed estimator is 
capable of recovering the $9$ (out of $14$ total) 
center frequencies that are being utilized for transmission. It accurately recovers the power
levels in the surveyed area with a smooth extrapolation to
zones were there are no measurements, and suggests possible
locations for the transmitters~\cite{rf_splines}.}
\vspace{-0.8cm}  \label{fig:psd_map}
\end{figure}

Sparse total LS variants are also available to cope with
uncertainty in the regression matrix, arising due to inaccurate channel estimation and
grid-mismatch effects~\cite{tut_rf_cartog}. Nonparametric spline-based PSD map 
estimators~\cite{rf_splines} have been also shown effective in capturing general propagation
characteristics including both shadowing and fading; see also Fig. \ref{fig:psd_map} for
an actual PSD atlas spanning $14$ frequency sub-bands.

% % % % % % % % % % % % % % % % % % % % % % % % % % % % % % % % % % % % % % % %
%                         Section V                                           %
% % % % % % % % % % % % % % % % % % % % % % % % % % % % % % % % % % % % % % % %

\section{Concluding remarks}
\label{sec:conc}

In this tutorial, the concept of dynamic network cartography is introduced
as a framework to construct maps of the dynamically evolving network state, 
in an efficient and scalable manner even for large-scale heterogeneous networks.
Focus is placed on key tasks geared to obtaining full yet succinct
representation of network state metrics such as link traffic and
path delays, as well as prompt and accurate identification of
network anomalies from possibly partial and corrupted measurement data.

Looking forward, the unceasing demand for continuous situational awareness calls
for innovative and large-scale distributed SP algorithms,
complemented by collaborative and adaptive monitoring platforms to
accomplish the objectives of network management and control.
Avenues where significant impact can be made include: i) 
judicious design of
critical cognition infrastructure to sense, learn, and adapt to the
environment where networks operate;
ii) development of scalable tools for distilling, summarizing, and tracking the
network state for the purpose of network management; iii) ensuring
robustness in the face of missing and grossly-corrupted
network data, in addition to possibly malicious
attacks; and iv) developing effective network adaptation techniques based
on global network inference, further impacting protocol designs, network
taxonomy, and categorization.

% % % % % % % % % % % % % % % % % % % % % % % % % % % % % % % % % % % % % % % %
%                         References                                          %
% % % % % % % % % % % % % % % % % % % % % % % % % % % % % % % % % % % % % % % %

%\newpage
\bibliographystyle{IEEEtranS}
\bibliography{IEEEabrv,biblio}

% Generated by IEEEtranS.bst, version: 1.13 (2008/09/30)
\begin{thebibliography}{10}
\providecommand{\url}[1]{#1}
\csname url@samestyle\endcsname
\providecommand{\newblock}{\relax}
\providecommand{\bibinfo}[2]{#2}
\providecommand{\BIBentrySTDinterwordspacing}{\spaceskip=0pt\relax}
\providecommand{\BIBentryALTinterwordstretchfactor}{4}
\providecommand{\BIBentryALTinterwordspacing}{\spaceskip=\fontdimen2\font plus
\BIBentryALTinterwordstretchfactor\fontdimen3\font minus
  \fontdimen4\font\relax}
\providecommand{\BIBforeignlanguage}[2]{{%
\expandafter\ifx\csname l@#1\endcsname\relax
\typeout{** WARNING: IEEEtranS.bst: No hyphenation pattern has been}%
\typeout{** loaded for the language `#1'. Using the pattern for}%
\typeout{** the default language instead.}%
\else
\language=\csname l@#1\endcsname
\fi
#2}}
\providecommand{\BIBdecl}{\relax}
\BIBdecl

\bibitem{Internet2}
\BIBentryALTinterwordspacing
 [Online]. Available: \url{http://www.internet2.edu}
\BIBentrySTDinterwordspacing

\bibitem{amp}
\BIBentryALTinterwordspacing
 [Online]. Available:
  \url{http://erg.cs.waikato.ac.nz/amp/matrix.php/ipv4/latency/NZ}
\BIBentrySTDinterwordspacing

\bibitem{ajwak10}
A.~Abdelkefi, Y.~Jiang, W.~Wang, A.~Aslebo, and O.~Kvittem, ``Robust traffic
  anomaly detection with principal component pursuit,'' in \emph{Proc. of the
  ACM CoNEXT Student Workshop}, Philadelphia, PA, Nov. 2010.

\bibitem{onlinetracking_bolzano10}
L.~Balzano, R.~Nowak, and B.~Recht, ``Online identification and tracking of
  subspaces from highly incomplete information,'' in \emph{Proc. of Allerton
  Conf. on Communication, Control, and Computing}, Monticello, IL, 2010.

\bibitem{Jayasumana}
V.~W. Bandara and A.~P. Jayasumana, ``Extracting baseline patterns in internet
  traffic using robust principal components,'' in \emph{Proc. IEEE Intl. Conf.
  on Local Computer Netw.}, Bonn, Germany, 2011.

\bibitem{bazerque}
J.~A. Bazerque and G.~B. Giannakis, ``Distributed spectrum sensing for
  cognitive radio networks by exploiting sparsity,'' \emph{IEEE Trans. Signal
  Process.}, vol.~58, pp. 1847--1862, Mar. 2010.

\bibitem{rf_splines}
J.~A. Bazerque, G.~Mateos, and G.~B. Giannakis, ``Group {L}asso on splines for
  spectrum cartography,'' \emph{IEEE Trans. Signal Process.}, vol.~59, pp.
  4648--4663, Oct. 2011.

\bibitem{Be2006Mreg}
M.~Belkin, P.~Niyogi, and V.~Sindhwani, ``Manifold regularization: A geometric
  framework for learning from labeled and unlabeled examples,'' \emph{J. Mach.
  Learn. Res.}, vol.~7, pp. 2399--2434, Dec. 2006.

\bibitem{Bertsekas_Book_Distr}
D.~P. Bertsekas and J.~N. Tsitsiklis, \emph{Parallel and Distributed
  Computation: Numerical Methods}.\hskip 1em plus 0.5em minus 0.4em\relax
  Athena-Scientific, 1999.

\bibitem{CLMW09}
E.~J. Candes, X.~Li, Y.~Ma, and J.~Wright, ``Robust principal component
  analysis?'' \emph{Journal of the ACM}, vol.~58, no.~1, pp. 1--37, 2011.

\bibitem{CT05}
E.~J. Candes and T.~Tao, ``Decoding by linear programming,'' \emph{IEEE Trans.
  Info. Theory}, vol.~51, no.~12, pp. 4203--4215, 2005.

\bibitem{candes_moisy_mc}
E.~Candes and Y.~Plan, ``Matrix completion with noise,'' \emph{Proc. of the
  IEEE}, vol.~98, pp. 925--936, 2009.

\bibitem{tomo_nowak}
R.~Castro, M.~Coates, G.~Liang, R.~Nowak, and B.~Yu, ``Network tomography:
  Recent developments,'' \emph{Statist. Sci.}, vol.~19, no.~3, pp. 499--517,
  2004.

\bibitem{CSPW11}
V.~Chandrasekaran, S.~Sanghavi, P.~R. Parrilo, and A.~S. Willsky,
  ``Rank-sparsity incoherence for matrix decomposition,'' \emph{SIAM J.
  Optim.}, vol.~21, no.~2, pp. 572--596, 2011.

\bibitem{Vaswani_Allerton_11}
Q.~Chenlu and N.~Vaswani, ``Recursive sparse recovery in large but correlated
  noise,'' in \emph{Proc. of Allerton Conf. on Communication, Control, and
  Computing}, Monticello, IL, 2011.

\bibitem{petrels_chi12}
Y.~Chi, Y.~C. Eldar, and R.~Calderbank, ``Petrels: Subspace estimation and
  tracking from partial observations,'' in \emph{Proc. of IEEE International
  Conference on Acoustics, Speech and Signal Processing}, Kyoto, Japan, Mar.
  2012.

\bibitem{nk}
D.~Chua, E.~Kolaczyk, and M.~Crovella, ``Network kriging,'' \emph{IEEE J. Sel.
  Areas Commun.}, vol.~24, 2006.

\bibitem{coates}
M.~Coates, Y.~Pointurier, and M.~Rabbat, ``Compressed network monitoring for
  {IP} and all-optical networks,'' in \emph{Proc. ACM Internet Measurement
  Conf.}, San Diego, CA, Oct. 2007.

\bibitem{conti}
G.~Conti, \emph{Security Data Visualization: Graphical Techniques for Network
  Analysis}.\hskip 1em plus 0.5em minus 0.4em\relax No Starch Press, 2007.

\bibitem{Cor09}
J.~Cort\'{e}s, ``Distributed {K}riged {K}alman filter for spatial estimation,''
  vol.~54, no.~12, pp. 2816--2827, Dec. 2009.

\bibitem{cressie}
N.~Cressie, ``The origins of kriging,'' \emph{Mathematical Geology}, vol.~22,
  no.~3, pp. 239--252, 1990.

\bibitem{vivaldi}
F.~Dabek, R.~Cox, F.~Kaashoek, and R.~Morris, ``Vivaldi: A decentralized
  network coordinate system,'' in \emph{Proc. of ACM SIGCOMM}, Portland, OR,
  Aug. 2004.

\bibitem{Do06CS}
D.~L. Donoho, ``Compressed sensing,'' \emph{IEEE Trans. Info. Theory}, vol.~52,
  no.~4, pp. 1289 --1306, Apr. 2006.

\bibitem{high-rank}
B.~Eriksson, L.~Balzano, and R.~Nowak, ``High-rank matrix completion,'' in
  \emph{Proc. of Intl. Conf. on Artificial Intell. and Stat.}, La Palma, Canary
  Islands, Apr. 2012.

\bibitem{F02}
M.~Fazel, ``Matrix rank minimization with applications,'' Ph.D. dissertation,
  Electrical Eng. Dept., Stanford University, 2002.

\bibitem{pedrocip12}
P.~A. Forero, K.~Rajawat, and G.~B. Giannakis, ``Semi-supervised dictionary
  learning for network-wide link load prediction,'' in \emph{Proc. Cognitive
  Information Processing Workshop}, Baiona, Spain, May 2012.

\bibitem{Fri07pwise}
J.~Friedman, T.~Hastie, H.~H\"{o}fling, and R.~Tibshirani, ``Pathwise
  coordinate optimization,'' \emph{The Annals of Applied Statistics}, vol.~1,
  no.~2, pp. 302--332, 2007.

\bibitem{increm_grad_grasmannian_bolzano12}
J.~He, L.~Balzano, and A.~Szlam, ``Incremental gradient on the {G}rassmannian
  for online foreground and background separation in subsampled video,'' in
  \emph{Proc. of IEEE Conference on Computer Vision and Pattern Recognition},
  Providence, Rhode Island, Jun. 2012.

\bibitem{tut_rf_cartog}
S.-J. Kim, E.~Dall'Anese, J.~A. Bazerque, K.~Rajawat, and G.~B. Giannakis,
  ``Advances in spectrum sensing and cross-layer design for cognitive radio
  networks,'' \emph{Elsevier, E-Reference Signal Processing}, 2012.

\bibitem{kim}
S.-J. Kim, E.~Dall'Anese, and G.~B. Giannakis, ``Cooperative spectrum sensing
  for cognitive radios using {K}riged {K}alman filtering,'' \emph{IEEE Jrnl.
  Sel. Topics in Signal Process.}, vol.~5, no.~1, pp. 24--36, Feb. 2011.

\bibitem{crowdad}
T.~King, S.~Kopf, T.~Haenselmann, C.~Lubberger, and W.~Effelsberg, ``{CRAWDAD}
  data set mannheim/compass (v. 2008-04-11),'' Downloaded from
  http://crawdad.cs.dartmouth.edu/mannheim/compass, Apr. 2008.

\bibitem{Kolaczyk_book}
E.~D. Kolaczyk, \emph{Statistical Analysis of Network Data: Methods and
  Models}.\hskip 1em plus 0.5em minus 0.4em\relax Springer, 2009.

\bibitem{lakhina}
A.~Lakhina, M.~Crovella, and C.~Diot, ``Diagnosing network-wide traffic
  anomalies,'' in \emph{Proc. of ACM SIGCOMM}, Portland, OR, Aug. 2004.

\bibitem{tiv}
S.~Lee, Z.~Zhang, S.~Sahu, and D.~Saha, ``On suitability of {E}uclidean
  embedding {I}nternet hosts,'' in \emph{SIGMETRICS}, Saint Malo, France, Jun.
  2006.

\bibitem{latencyprediction}
Y.~Liao, P.~Geurts, and G.~Leduc, ``Network distance prediction based on
  decentralized matrix factorization,'' in \emph{Proc. of IFIP Networking
  Conf.}, Chennai, India, May 2010.

\bibitem{mairal}
J.~Mairal, J.~Bach, J.~Ponce, and G.~Sapiro, ``Online learning for matrix
  factorization and sparse coding,'' \emph{Jrnl. of Machine Learning Research},
  vol.~11, pp. 19--60, Jan. 2010.

\bibitem{dyn_anomal}
M.~Mardani, G.~Mateos, and G.~B. Giannakis, ``Dynamic anomalography: Tracking
  network anomalies via sparsity and low rank,'' \emph{IEEE Jrnl. Sel. Topics
  in Signal Process.}, 2012, see also arXiv:1208.4043v1 [cs.NI].

\bibitem{tsp_rankminimization_2012}
------, ``In-network sparsity-regularized rank minimization: Applications and
  algorithms,'' \emph{IEEE Trans. Signal Process.}, 2012, see also
  arXiv:1203.1507v1 [cs.MA].

\bibitem{tit_recovery_2012}
------, ``Recovery of low-rank plus compressed sparse matrices with application
  to unveiling traffic anomalies,'' \emph{IEEE Trans. Info. Theory}, 2012, see
  also arXiv:1204.6537v1 [cs.IT].

\bibitem{MGRA98}
K.~V. Mardia, C.~Goodall, E.~J. Redfern, and F.~J. Alonso, ``The {K}riged
  {K}alman filter,'' \emph{Test}, vol.~7, no.~2, pp. 217--285, Dec. 1998.

\bibitem{myers76}
K.~Myers and B.~Tapley, ``Adaptive sequential estimation with unknown noise
  statistics,'' \emph{{IEEE} Trans. Automat. Contr.}, vol.~21, no.~4, pp.
  520--523, Aug. 1976.

\bibitem{nemh}
G.~L. Nemhauser, L.~A. Wolsey, and M.~L. Fisher, ``An analysis of
  approximations for maximizing submodular set functions - {I},''
  \emph{Mathematical Programming}, no.~1, pp. 265--294, Dec. 1978.

\bibitem{rrd}
\BIBentryALTinterwordspacing
T.~Oetiker. About rrdtool. [Online]. Available:
  \url{http://people.ee.ethz.ch/~oetiker/webtools/rrdtool}
\BIBentrySTDinterwordspacing

\bibitem{Ra2007SLT}
R.~Raina, A.~Battle, H.~Lee, B.~Packer, and A.~Y. Ng, ``Self-taught learning:
  transfer learning from unlabeled data,'' in \emph{Proceedings of the 24th
  Intl. Conf. on Machine learning}, ser. ICML '07, 2007, pp. 759--766.

\bibitem{rajassp12}
K.~Rajawat, E.~Dall'Anese, and G.~B. Giannakis, ``Dynamic network kriging,'' in
  \emph{Proc. {IEEE} Statistical Signal Processing Workshop}, Ann Arbor, MI,
  Aug. 2012, see also arXiv:1204.5507v1 [cs.NI].

\bibitem{Recht_Parallel_2011}
B.~Recht and C.~Re, ``Parallel stochastic gradient algorithms for large-scale
  matrix completion,'' 2011, (submitted).

\bibitem{Roughan}
M.~Roughan, ``A case study of the accuracy of {SNMP} measurements,''
  \emph{Journal of Electrical and Computer Engineering}, vol. 2010, 2010,
  article ID 812979.

\bibitem{Shavitt}
Y.~Shavitt, X.~Sun, A.~Wool, and B.~Yener, ``Computing the unmeasured: An
  algebraic approach to internet mapping,'' in \emph{Proc. IEEE Intl. Conf. on
  Computer Commun.}, Anchorage, Alaska, Apr. 2001.

\bibitem{soule}
A.~Soule, A.~Lakhina, N.~Taft, K.~Papagiannaki, K.~Salamatian, A.~Nucci,
  M.~Crovella, and C.~Diot, ``Traffic matrices: Balancing measurements,
  inference and modeling,'' in \emph{Proc. ACM SIGMETRICS}, Banff, AB, Jun.
  2005.

\bibitem{soulekalman}
A.~Soule, K.~Salamatian, A.~Nucci, and N.~Taft, ``Traffic matrix tracking using
  kalman filters,'' \emph{SIGMETRICS Perform. Eval. Rev.}, vol.~33, no.~3, pp.
  24--31, Dec. 2005.

\bibitem{Ti1996lasso}
R.~Tibshirani, ``Regression shrinkage and selection via the {L}asso,''
  \emph{Journal of the Royal Statistical Society. Series B (Methodological)},
  vol.~58, no.~1, pp. 267--288, 1996.

\bibitem{Dictionary_learning_SP_mag_10}
I.~To\v{s}i\'{c} and P.~Frossard, ``Dictionary learning,'' \emph{IEEE Signal
  Process. Mag.}, vol.~28, pp. 27--38, Mar. 2010.

\bibitem{Wu11}
X.~Wu, K.~Yu, and X.~Wang, ``On the growth of internet application flows: A
  complex network perspective,'' in \emph{Proc. IEEE Intl. Conf. on Computer
  Commun.}, Shangai, China, Jun. 2011.

\bibitem{Zha05est}
Y.~Zhang, M.~Roughan, C.~Lund, and D.~L. Donoho, ``Estimating point-to-point
  and point-to-multipoint traffic matrices: an information-theoretic
  approach,'' \emph{IEEE/ACM Transactions on Networking}, vol.~13, no.~5, pp.
  947 -- 960, Oct. 2005.

\bibitem{zggr05}
Y.~Zhang, Z.~Ge, A.~Greenberg, and M.~Roughan, ``Network anomography,'' in
  \emph{Proc. ACM SIGCOM Conf. on Interent Measurements}, Berkeley, CA, Oct.
  2005.

\bibitem{zrwq09}
Y.~Zhang, M.~Roughan, W.~Willinger, and L.~Qiu, ``Spatio-temporal compressive
  sensing and internet traffic matrices,'' in \emph{Proc. of ACM SIGCOM Conf.
  on Data Commun.}, New York, USA, Oct. 2009.

\end{thebibliography}

\end{document}